\documentclass{article}
\usepackage{ijcai25}

\usepackage{times}
\usepackage{soul}
\usepackage{url}
\usepackage[hidelinks]{hyperref}
\usepackage[utf8]{inputenc}
\usepackage[small]{caption}
\usepackage{graphicx}
\usepackage{amsmath}
\usepackage{amsthm}
\usepackage{booktabs}
\usepackage{algorithm}
\usepackage{algorithmic}
\usepackage[switch]{lineno}

\usepackage{bm}
\usepackage{amssymb}
\usepackage{amsmath}
\usepackage{makecell}
\usepackage{multirow}
\usepackage{subcaption}
\usepackage{hyperref}

\title{Scalable Safe Multi-Agent Reinforcement Learning for Multi-Agent System}

	\author{
Haikuo Du
\and
Fandi Gou \And
Yunze Cai$^{*}$\\
\affiliations
Department of Automation, Shanghai Jiao Tong University\\
\emails
\{duhaikuo1013, finley-gou3, yzcai\}@sjtu.edu.cn
}

\begin{document}

\maketitle
\begin{abstract}
	Safety and scalability are two critical challenges faced by practical Multi-Agent Systems (MAS). However, existing Multi-Agent Reinforcement Learning (MARL) algorithms that rely solely on reward shaping are ineffective in ensuring safety, and their scalability is rather limited due to the fixed-size network output.  To address these issues, we propose a novel framework, Scalable Safe MARL (SS-MARL), to enhance the safety and scalability of MARL methods. Leveraging the inherent graph structure of MAS, we design a multi-layer message passing network to aggregate local observations and communications of varying sizes. Furthermore, we develop a constrained joint policy optimization method in the setting of local observation to improve safety. Simulation experiments demonstrate that SS-MARL achieves a better trade-off between optimality and safety compared to baselines, and its scalability significantly outperforms the latest methods in scenarios with a large number of agents.
\end{abstract}

\section{Introduction}
Reinforcement learning (RL) techniques have experienced considerable advancement in recent years. However, the transition of these techniques from virtual environments to real-world physical applications still presents numerous challenges. Among these challenges, safety remains a critical concern, particularly in fields such as robotics  \cite{REVIEW[robotics]}, autonomous driving \cite{REVIEW[autonomous_driving]}, multi-agent systems (MAS) \cite{MACPO}, and even the fine-tuning of large language models  \cite{REVIEW_safe_RLHF}. In the context of MAS, Multi-Agent Reinforcement Learning (MARL) demonstrates significant advantages in addressing complex cooperative tasks. During the execution of these tasks, agents must adhere to a variety of both local and global safety constraints. These safety constraints are designed to mitigate risks to the agents themselves as well as to other entities within the environment. Therefore, it is essential to study safe decision-making under these safe constraints within MARL frameworks.
\par 
Most studies in the field of MAS based on MARL treat safety constraints as negative penalty in rewards \cite{DRL_Obstacle_Avoidance_1,DRL_Obstacle_Avoidance_2,DRL_Obstacle_Avoidance_3}. However, because negative penalties can conflict with other positive rewards, neither the final policy nor the optimal policy inherently guarantees the satisfaction of safety constraints \cite{REVIEW[robotics]}. This can result in numerous states that violate safety constraints, which are unacceptable in real-world applications. To address these challenges and inspired by safe RL algorithms on single agent  \cite{QMIX,CPO,PPO_Larg_TPRO_Larg}, some researchers developed a series of safe MARL algorithms \cite{CMIX,MACPO}. Among these algorithms, safety constraints are modeled as cost constraints, rather than negative penalty in rewards. These algorithms are designed to maintain safety throughout both the training and testing phases by taking cost constraints into account in policy optimization. 
\par 
Despite advancements in safe MARL algorithms that enhance safety, the challenge of exponential state space growth in MAS remains significant as the number of agents scales up. To tackle the exponential explosion problem, research on scalability in MARL aims to develop effective algorithms that can tranfer from small-scale training scenarios to large-scale testing scenarios. Some researchers \cite{INFORMARL,TEM[IJCAI]} have utilized the Centralized Training Decentralized Execution (CTDE) framework to achieve scalability in the setting of local observation. However, the issue with these methods lies in how better to utilize the local observations and communications of agents to achieve task objectives, including both optimality and safety objectives, especially when transferred to large-scale tasks.
\par
To deal with the aforementioned issues, this paper proposes a novel framework called Scalable Safe Multi-Agent Reinforcement Learning (SS-MARL). SS-MARL has the following features: (1) It utilizes graph neural networks (GNNs) to achieve implicit communication between agents, while enhancing sampling efficiency during the training phase; (2) It employs constrained joint policy optimization, which can handle multiple constraints to ensure safety during both the training and testing phases; (3) It is capable of zero-shot transferring models trained on small-scale tasks to large-scale tasks while maintaining high safety level. 
\par
We evaluated SS-MARL in Multi-agent Particle Environment (MPE) \cite{MADDPG}. Meanwhile, we improved MPE to Safe MPE for safe MARL algorithms. Whether in the training or testing phases, the experimental results of SS-MARL demonstrate superior performance compared to other latest algorithms. Due to page limitations, experiments on more cooperative tasks and the hardware implementation are shown in the appendix.

\section{Related Works}
\subsection{Safe MARL}
A research paradigm of safe RL is Constrained Markov Decision Processes (CMDPs), which take the cost constraints into account during MDPs state transitions. Following this paradigm, algorithms such as CPO \cite{CPO}, TRPO-Lagrangian and PPO-Lagrangian \cite{PPO_Larg_TPRO_Larg} were proposed, all of which are inspired by TRPO \cite{TRPO}. When it comes to multi-agent settings, each agent must not only adhere to its own cost constraints but also ensure that the joint behaviors of all agents possess a safety guarantee while maximizing the total reward. To deal with these issues, CMIX \cite{CMIX} was proposed based on QMIX \cite{QMIX}, which modifies the reward function to address peak cost constraints, but it doesn't provide a convergence proof. Inspired by HATRPO \cite{HATRPO}, MACPO and MAPPO-Lagrangian \cite{MACPO} proved the theoretical guarantees of both monotonic improvement in reward and satisfaction of cost constraints, under the premise that the state value functions are known. Unfortunately, these algorithms do not consider the local observations of agents, nor do they provide solutions when agents have multiple constraints.
\subsection{Partial Observability in MARL}
In real-world MAS, agents can only observe partial information about the environment.  The research paradigm that takes the partial observability into account is called Decentralized Partially Observable Markov Decision Processes (Dec-POMDPs)  \cite{DecPOMDP}. In Dec-POMDPs, partial observability leads agents to treat other unobservable agents as part of the environment. This results in the non-stationarity of the environment \cite{REVIEW[non-stationarity]} and violates the Markov property. \par
In the context of MAS engaged in cooperative tasks, information sharing is important to tackle the challenge of partial observability \cite{REVIEW[communication]}. Agents share local observations with other agents, which mitigates the impact of partial observability. CommNet \cite{CommNet} is a pioneering work that introduced communication into MARL by employing a shared communication neural network. However, it assumes that all agents can communicate with each other pairwise which is impractical. GAXNet \cite{GAXNet} introduces an attention mechanism that allows for weighted selection of messages from other agents, but it is limited by the requirement to fix the maximum number of agents before training. Addressing this issue, TEM \cite{TEM[IJCAI]} proposes a Transformer-based Email mechanism for scalable inter-agent communication. Nevertheless, the TEM approach only uses inter-agent communication to complete simple task and ignores the constraints in MAS.

\subsection{Scalability in MARL}
The CTDE framework in MARL raises a compelling question: Can models trained on small-scale tasks be transferred to larger-scale tasks, given the agents' decentralized policy execution? Most MARL algorithms, such as MADDPG \cite{MADDPG}, MAPPO \cite{MAPPO}, cannot achieve this goal because their actor networks' input sizes are entirely dependent on the state dimensions during training. With the recent success of Graph Neural Networks (GNNs), researchers have increasingly recognized the potential of integrating GNNs with the inherent graph structure in MAS to tackle this scalability issue in MARL.\par 
DGN \cite{DGN} utilizes multi-head attention mechanism and Graph Convolutional Networks (GCNs) \cite{GCN} to achieve scalability. However, it poses an oversimplified hypothesis that agents only communicate with the nearest three agents, encountering significant limitations in practical applications. EMP \cite{EMP} also employs a GNNs approach based on the distance between entities, but it assumes that all entity states are known at the beginning of an episode, failing to effectively handle situations where obstacles are not observed initially but appear later. InforMARL \cite{INFORMARL} integrates UniMP \cite{UniMP} networks into actor and critic networks, proposing a scalable MARL framework capable of selecting weights for local observations or communication through an attention mechanism. However, its drawback lies in the continued use of reward penalty to maintain safety, which results in the underutilization of aggregated information by the GNNs.
\begin{figure*}[h]
	\centering
	\includegraphics[width=0.90\linewidth]{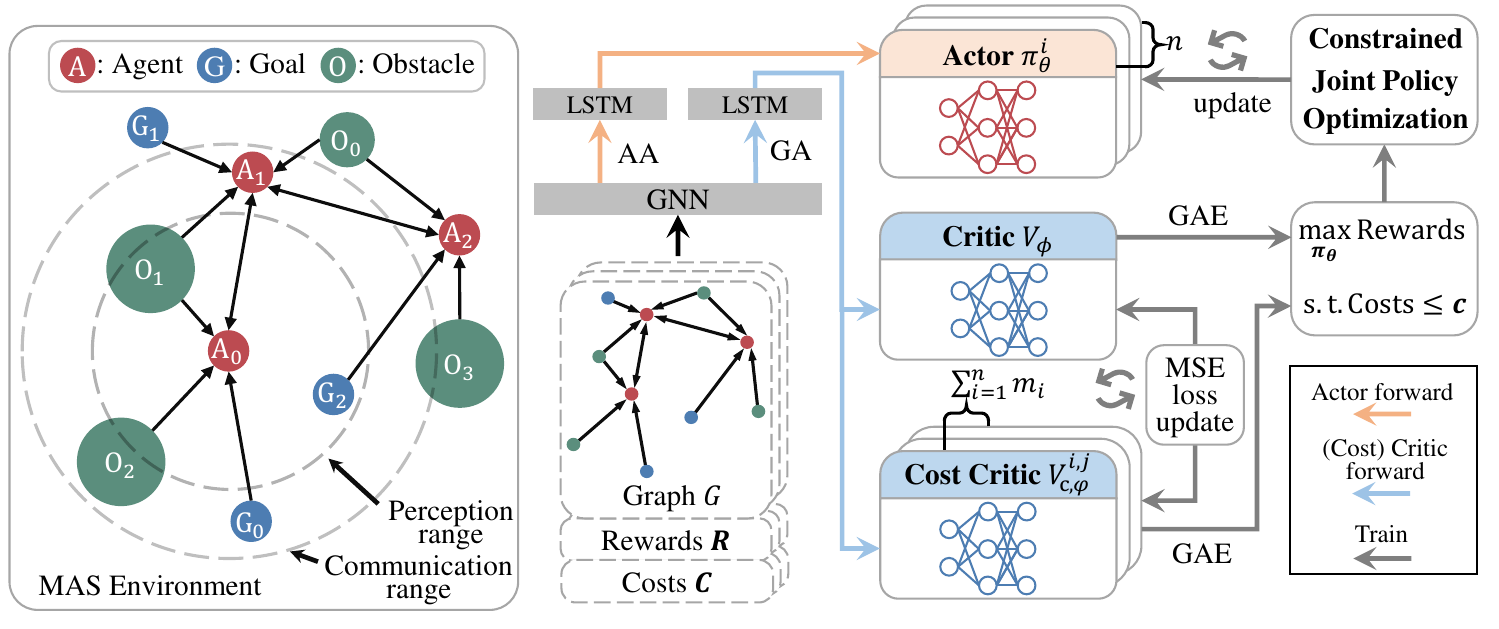}
	\caption{Overview of SS-MARL. AA: Agent Aggregation, GA: Graph Aggregation, GAE: Generalized Advantage Estimation.}
	\label{fig:framework}
\end{figure*}
\section{Problem Formulation}
\label{section:pf}
We formulate the safe MARL problem as a Constrained Markov Game (CMG) $\left \langle \bm{N},S,\bm{A}, P,\rho_0,R,\gamma,\bm{C},\bm{c},\gamma_{\text{c}} \right \rangle$, where $\bm{N}=\{1,2,...,n\}$ is the set of agents, $S$ and $\bm{A}=\prod_{i=1}^{n}A_i$ stand for the state space and the joint action space respectively, $P:S\times \bm{A}\times S \to \mathbb{R}$ is the state transition function, $\rho_0$ is the initial state distribution, $R=S\times \bm{A}\to \mathbb{R}$ is the joint reward function, $\bm{C}=\left \{ C_i^j:S\times A_i\to \mathbb{R}|i\in \bm{N},1\le j\le m_i  \right \} $ is the set of cost functions, each agent $i$ has $m_i$ cost functions, $\bm{c}=\left \{ c_i^j:\mathbb{R}|i\in \bm{N},1\le j\le m_i  \right \}$ is the set of corresponding cost-constraining values, $\gamma,\gamma_{\text{c}}\in \left[0,1\right)$ represent the discount factor of rewards and costs respectively. At time step \( t \), agents are in state \( S^t \), and agent \( i \) takes action \( A_i^t \) according to its policy \( \pi_i (A_i^t | S^t) \), leading to a joint policy \( \bm{\pi}(\bm{A}^t|S^t) = \prod_{i=1}^{n} \pi_i (A_i^t | S^t) \). \par
In this paper, we consider a fully cooperative setting where all agents share the same reward function. The objective in this case is to maximize the total reward while also trying to satisfy each agent's safety constraints.
\begin{align}
	\max_{\bm{\pi}}J(\bm{\pi}) &\triangleq \mathbb{E}_{S^{0} \sim \rho_0, 	\bm{A}^{0:\infty} \sim \bm{\pi} } \left[ \sum_{t=0}^{\infty} \gamma^t R(S^t,\bm{A}^t) \right], \nonumber \\
	\text{s.t. }  J_{\text{c},i}^j(\bm{\pi}) &\triangleq \mathbb{E}_{S^{0} \sim \rho_0, 	A^{0:\infty}_i \sim \pi_i } \left[ \sum_{t=0}^{\infty} \gamma_{\text{c}}^t C_i^j(S^t,A^t_i) \right] \le c_i^j, \nonumber \\
	&\forall i\in \bm{N},1\le j\le m_i.
	\label{equ1}
\end{align}

The joint policies that satisfy the constraints in Equation \ref{equ1} are referred to as \textbf{feasible}. The following definitions are based on the reward function, the cost version definitions are similar\footnote{Variables of the cost version are denoted with the subscript 'c'.}.
We can define the multi-agent advantage ($\mathcal{A}$) function based on the multi-agent state-action value ($Q$) function. Here, \( i_{1:h} \) and \( j_{1:k} \) are two disjoint subsets of the set \( \bm{N} \), and \( -i_{1:h} \) denotes the complement of \( i_{1:h} \) with respect to \( \bm{N} \).
\begin{align}
	\mathcal{A}_{\bm{\pi}}^{i_{1: h}}\left(S, \bm{A}_{j_{1: k}}, \bm{A}_{i_{1: h}}\right) & \triangleq   Q_{\bm{\pi}}^{j_{1: k} \cup i_{1: h}}\left(S, \bm{A}_{j_{1: k}} \cup \bm{A}_{i_{1: h}}\right) \nonumber \\ & - Q_{\bm{\pi}}^{j_{1: k}}\left(S, \bm{A}_{j_{1: k}}\right). 
\end{align}

Subsequently, we can define the joint surrogate return, where \( \bm{\pi}, \hat{\bm{\pi}}, \bar{\bm{\pi}} \) represent three different joint policies.
\begin{align}
	L_{\bm{\pi}}^{i_{1:h}}\left(\hat{\bm{\pi}}_{i_{1:h-1}},\bar{\pi}_{i_{h}}\right) & \triangleq  \mathbb{E}_{S\sim\rho_0,\bm{A}_{i_{1:h-1}}\sim\hat{\bm{\pi}}_{i_{1:h-1}},A_{i_{h}}\sim\bar{\pi}_{i_{h}}} \nonumber \\ & \left[\mathcal{A}_{\bm{\pi}}^{i_{h}} \left(S, \bm{A}_{i_{1: h-1}}, A_{i_{h}}\right)\right].
\end{align}

We assume that the joint policy after $k$ updates is \( \bm{\pi}^k = \{\pi^k_1, \pi^k_2,...,\pi^k_n\} \), and $\bm{\pi}^k_{i_{1:h}}$ represents the subset of $\bm{\pi}^k$ for which the subscripts are in $i_{1:h}$. The joint policy can be improved if each agent in $i_{1:h}$ sequentially solves the following optimization problem:
\begin{align}
	\pi^{k+1}_{i_h} &= \arg\max_{\pi} L_{\bm{\pi}^{k}}^{i_{1: h}}\left(\bm{\pi}^{k+1}_{i_{1: h-1}}, \pi\right)-\nu D_{\mathrm{KL}}^{\mathrm{max}}(\pi, \pi_{i_h}^k), \nonumber \\
	&\text{s.t.   } D_{\text{KL}}^{\text{max}}\left(\pi^{k}_{i_h}, \pi\right) \leq \delta_{i_h},\nonumber \\
	&J_{\text{c},i_{h}}^{j}\left(\bm{\pi}^{k}_{i_{1:h-1}}\cup \pi\right)\leq c^{j}_{i_{h}} , \forall j = 1, \ldots, m_{i_{h}}.
	\label{opt1}
\end{align}

Inspired by \cite{MACPO} and \cite{HATRPO}, it can be proven that the joint policy will be improved monotonically while satisfying the cost constraints by following the sequential update process in Equation \ref{opt1}. Detailed proofs are given in the appendix.

\section{Approach}

The framework of SS-MARL is shown in Figure \ref{fig:framework}. We extract the inherent graph $G:(V,E)$ in MAS at time $t$. The vertices of the graph are all entities in the environment, including obstacles, agents, and corresponding goals. We use encoding to represent the type of vertices (0: obstacle, 1: agent, 2: goal). Every agent has a communication range and a perception range. Agents can perceive other agents and obstacles within their perception range, forming unidirectional edges from the perceived entities to themselves. Within their communication range, agents can communicate bidirectionally with other agents. Additionally, each agent can form a unidirectional connection edge with its goal according to its task. The detailed definitions of vertices and edges in graph 
$G$ are provided in Table \ref{tab:edge_type}.
\begin{table}[h]
	\centering
	\begin{tabular}{p{2.4cm}<{\centering}ccc}
		\toprule
		\makecell[c]{Edge \\ Type} & \makecell[c]{Source \\ Vertex} & \makecell[c]{Target \\ Vertex} & \makecell[c]{Edge \\ Features} \\
		\midrule
		
		\makecell[c]{Obstacle-Agent} & \makecell[c]{0} & \makecell[c]{1} & \makecell[c]{$[\mathbf{S}_\text{0} - \mathbf{S}_\text{1}]$} \\
		
		\makecell[c]{Goal-Agent} & \makecell[c]{2} & \makecell[c]{1} & \makecell[c]{$[\mathbf{S}_\text{2} - \mathbf{S}_\text{1}]$} \\
		
		\multirow[c]{2}{2.4cm}{Agent 0-Agent 1} & \makecell[c]{1(Agent 0)} & \makecell[c]{1(Agent 1)} & \makecell[c]{$[\mathbf{S}_\text{10} - \mathbf{S}_\text{11}]$} \\
		& \makecell[c]{1(Agent 1)} & \makecell[c]{1(Agent 0)} & \makecell[c]{$[\mathbf{S}_\text{11} - \mathbf{S}_\text{10}]$} \\
		\bottomrule
	\end{tabular}
	\caption{Definitions of edges and vertices. $\mathbf{S}$ represents the state of the entity. For the double integrator model, it consists of position and velocity.}
	\label{tab:edge_type}
\end{table}

In the original Dec-POMDPs setting, there is no communication between agents, and agents can only make decisions based on their local observations. However, in SS-MARL, due to the communication among agents, every agent can obtain more global information beyond their local observations by communicating with neighboring agents. In SS-MARL, we focus on solving the problem of how agents can efficiently obtain global information beyond local observations through communication with other agents.\par 
Besides, graph $G$ not only contains the global information of the environment but also includes the local observations of all agents and the communication topology. It is fed into the actor, critic, and cost critic, each of which outputs actions, reward state values, and cost state values, respectively. During the execution phase, actions are directly applied to the environment. In the training phase, critic and cost critic are used to guide constrained joint policy optimization. Meanwhile, the critic and the cost critic are updated via gradient descent using the mean squared error (MSE) loss of the temporal difference error, for rewards and costs respectively.

\subsection{Actor, Critic and Cost Critic}
Safe MARL algorithms extend the Actor-Critic architecture of MARL by incorporating an additional cost critic component. Cost critic is designed to estimate the cost state value functions. It plays a pivotal role in constrained joint policy optimization by serving as cost constraints, ensuring that the joint policy is optimized in a manner that maintains safety.\par
\begin{figure}[htbp]
	\centering
	\includegraphics[width=0.49\textwidth]{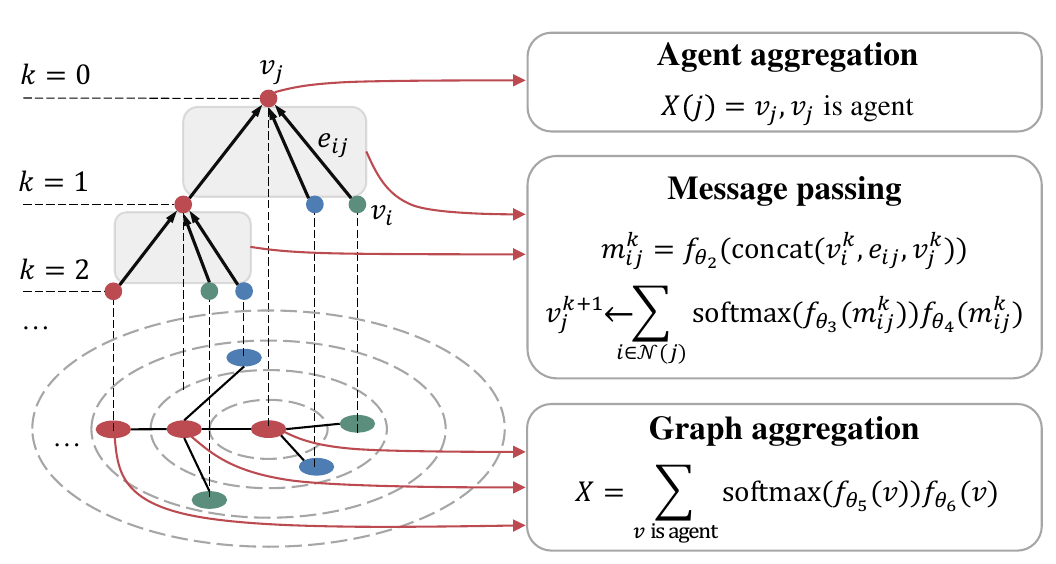}
	\caption{Visual Representation of message passing and aggregations in GNN. (Note: The embedding layer during the initial message passing is not shown in the above figure)}
	\label{fig:network}
\end{figure}

The actor needs to extract information from graph $G$ that includes both local observation and communication details of agents, while the critic and cost critic require global information from $G$. Here, we employ a graph attention mechanism-based message passing model to achieve this. In the GNN backbone, vertex features $V = \{v_i | 1 \le i \le |V|\}$ first pass through an embedding layer $f_{\theta_1}$ to generate embedding vectors. Then, for each connected edge $\{i, j\} $, the edge features $e_{ij} \in E$ and source vertex features are concatenated with target vertex features and processed through a MLP $f_{\theta_2}$ to generate a message, i.e., $m^k_{ij} = f_{\theta_2}(\text{concat}(f_{\theta_1}(v^k_i), e_{ij}, f_{\theta_1}(v^k_j)))$. Subsequently, an attention layer aggregates these messages: \( v^{k+1}_i = \sum_{j \in \mathcal{N}(i)} \text{softmax}(f_{\theta_3}(m^k_{ij})) \times f_{\theta_4}(m^k_{ij}) \). Here, \( \text{softmax}(f_{\theta_3}(m^k_{ij})) \) is referred to as the attention weight, a number between 0 and 1, representing the importance of the communication or observation between entity \( i \) and entity \( j \). The attention mechanism allows agents to select messages from connected edges based on their importance. We use multiple such message passing layers to enable agents to obtain information from more distant agents, mitigating the impact of partial observability. The specific message passing process is shown in Figure \ref{fig:network}. It is worth noting that the embedding layer $f_{\theta_1}$ is used to standardize vertex features dimension solely during the initial message passing and is not required for subsequent message passing.\par
The actor, critic and cost critic share the same GNN backbone, but they differ in the aggregation methods. After \( k \) message passing, graph $G$ becomes \( G^k: (V^k, E), V^k = \{v^k_i | 1 \le i \le |V|\} \). \( V^k_{\text{agent}} \subset V^k \) represents the subset of all agent vertices. The  actor performs \textbf{Agent Aggregation} (AA), which aggregates observation and communication related to a single agent. For agent \( i \), agent aggregation selects the vector of the vertex corresponding to agent \( i \) from \( V^k_{\text{agent}} \) as output. The critic and the cost critic perform \textbf{Graph Aggregation} (GA), which aggregates the vertices of all agents by passing \( V^k_{\text{agent}} \) through an attention layer, i.e., \( X = \sum_{v \in V^k_{\text{agent}}} \text{softmax}(f_{\theta_5}(v)) \times f_{\theta_6}(v) \). It is worth noting that, although GA aggregates more environment information than AA, both have output vectors of the same length. This length is independent of the number of agents, which also supports the scalability of SS-MARL. \par
After the aggregation process, the actor, critic and cost critic all require an LSTM \cite{LSTM} to address the challenges associated with violations of the Markov property in the environment. Subsequently, a diagonal Gaussian layer is utilized for the actor's stochastic output. For the critic and the cost critic, a MLP layer is engaged to compute and output the reward state value and the cost state value, respectively.

\subsection{Constrained Joint Policy Optimization}
As shown in Equation \ref{opt1}, it is feasible to update the joint policy with cost constraints. In practical implementation, we parameterize the joint policy \( \bm{\pi}^k = \prod_{i=1}^{n}\pi_{\theta_i^k} \) after $k$ updates and use average KL distance constraints derived from random sampling to approximate the maximum KL distance constraint. Thus, the optimization problem can be simplified to Equation \ref{opt2}. The multi-agent advantage functions for rewards and costs can be derived from the critic and the cost critic, respectively, using Generalized Advantage Estimation (GAE) \cite{GAE}.
\begin{align}
	&\theta^{k+1}_{i_h} = \arg\max_{\theta} \nonumber \\
	& \mathbb{E}_{S \sim \rho_0, \bm{A}_{i_{-h}} \sim \bm{\pi}^{k}_{i_{-h}}, A_{i_{h}} \sim \pi_{\theta}}\left[\mathcal{A}_{\bm{\pi}^{k}}^{i_{h}}\left(S, \bm{A}_{i_{-h}}, A_{i_{h}}\right)\right] \nonumber \\
	& \text{s.t. } J_{\text{c},i_h}^{j}\left(\bm{\pi}^{k}\right) + \mathbb{E}_{S \sim \rho_0, A_{i_{h}} \sim \pi_{\theta}}\left[\mathcal{A}_{\text{c},\bm{\pi}^{k}}^{i_h,j}\left(S, A_{i_{h}}\right)\right] \leq c^{j}_{i_{h}}, \nonumber \\
	&\forall j = 1, \ldots, m_{i_{h}}, \text{ and } \bar{D}_{\text{KL}}\left(\pi_{\theta^{k}_{i_h}}, \pi_{\theta}\right) \leq \delta.
	\label{opt2}
\end{align}

 Similar to \cite{HATRPO} and \cite{MACPO}, Equation \ref{opt2} can be approximated using a first-order Taylor expansion for the objective function and cost constraints, and a second-order approximation for the KL divergence constraint, as shown below: 
\begin{align}
	\theta^{k+1}_{i_h} &= \arg\max_{\theta}(\bm{G}_{i_h})(\theta-\theta^{k}_{i_h}) \nonumber \\
	& \text{s.t. } D_{i_h}^j + (\bm{B}_{i_h}^j)(\theta-\theta^{k}_{i_h})\le 0, \forall j = 1, \ldots, m_{i_{h}},\nonumber \\
	& \text{and } \frac{1}{2}(\theta-\theta^{k}_{i_h})^T\bm{E}_{i_h} (\theta-\theta^{k}_{i_h}) \leq \delta.
	\label{opt3}
\end{align}

Here, $\bm{G}_{i_h}$ is the gradient of the objective of agent $i_h$, $\bm{B}_{i_h}^j$ is the gradient of $j$th cost constraint of agent $i_h$, $\bm{E}_{i_h}$ is the Hessian matrix of the average KL distance of agent $i_h$. Subsequently, a trust-region based dual method can be employed to solve Equation \ref{opt3}. \par
However, the approximation may result in an infeasible policy $\bar{\theta}_{i_h}$. References \cite{CPO} and \cite{MACPO} address this issue only when $m_{i}=1, \forall i\in \bm{N}$. We now propose a solution for the case when $m_{i} \geq 1$. This involves a multi-objective optimization problem to reduce multiple cost values. To avoid calculating the complex Pareto front, we construct a weighted objective function, as shown below: 
	\begin{align}
	&\min_\theta \sum_{j=1}^{m_{i_h}}\left(\beta_j\bm{B}_{i_h}^{j}\right)(\theta-\bar{\theta}_{i_h}) \nonumber \\
	&\text{s.t.  } \frac{1}{2}(\theta-\bar{\theta}_{i_h})^T\bm{E}_{i_h} (\theta-\bar{\theta}_{i_h}) \leq \delta.
	\label{opt4}
\end{align}

The weight $\beta_j$ describes the importance of cost constraints. Constraints that significantly exceed their target values require higher weights, while those that meet the target values do not need reduction and should have a weight of zero.
\begin{equation}
	\resizebox{.9\linewidth}{!}{$
		\displaystyle
	\beta_j=\left\{\begin{matrix} 
		\frac{\text{exp}(P_j)}{\sum_{P_q>0}\text{exp}(P_q)}  \text{ , }P_j>0\\  
		0   \text{ , }P_j\le 0
	\end{matrix}\right. ,
	P_j = J_{\text{c},i_h}^{j}\left(\bm{\pi}_{\bar{\theta}_{i_h}}\right) - c_{i_h}^j.
$}
\end{equation}

To solve Equation \ref{opt4}, we can use a TRPO recovery step \cite{TRPO} on weighted matrix $\mathcal{B}_{i_h} =\sum_{j=1}^{m_{i_h}}\beta_j\bm{B}_{i_h}^{j}$ to recover the feasible policy, as shown below:
\begin{equation}
	\resizebox{.89\linewidth}{!}{$
		\displaystyle
		\theta^{k+1}_{i_h}=\theta^k_{i_h}- \alpha_j \sqrt{\frac{2\delta}{(\mathcal{B}_{i_h}) \left(\bm{E}_{i_h}\right)^{-1} (\mathcal{B}_{i_h})^T}}\left(\bm{E}_{i_h}\right)^{-1} (\mathcal{B}_{i_h})^T.
		$}
\end{equation}

$\alpha_j$ is adjusted through backtracking line search. Thus, even if the infeasible policy violates multiple cosntraints, it can still be recovered to a feasible policy by a recovery step.


\section{Experiments}
\begin{figure*}[htbp]
	\centering
	\begin{subfigure}[b]{0.38\textwidth}
		\includegraphics[width=\textwidth]{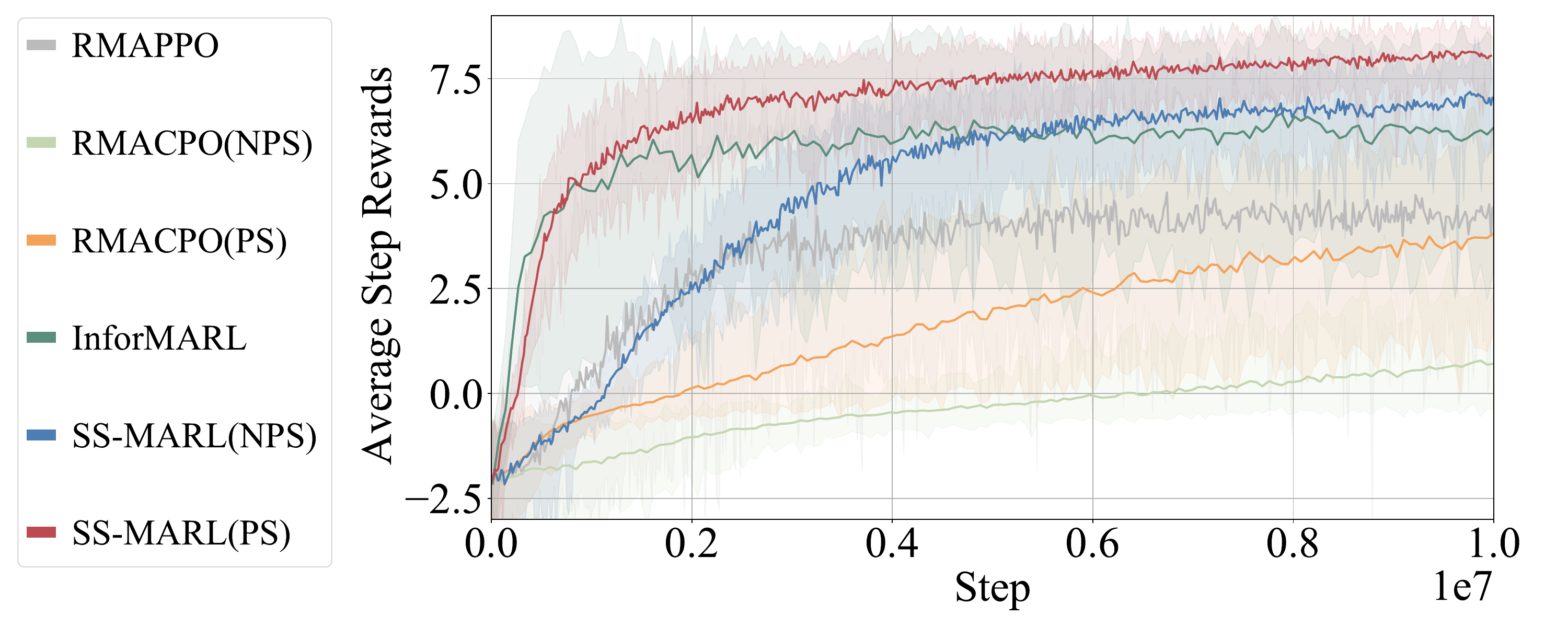}
		\caption{$n=3$}
		\label{fig:sub1}
	\end{subfigure}
	\begin{subfigure}[b]{0.30\textwidth}
		\includegraphics[width=\textwidth]{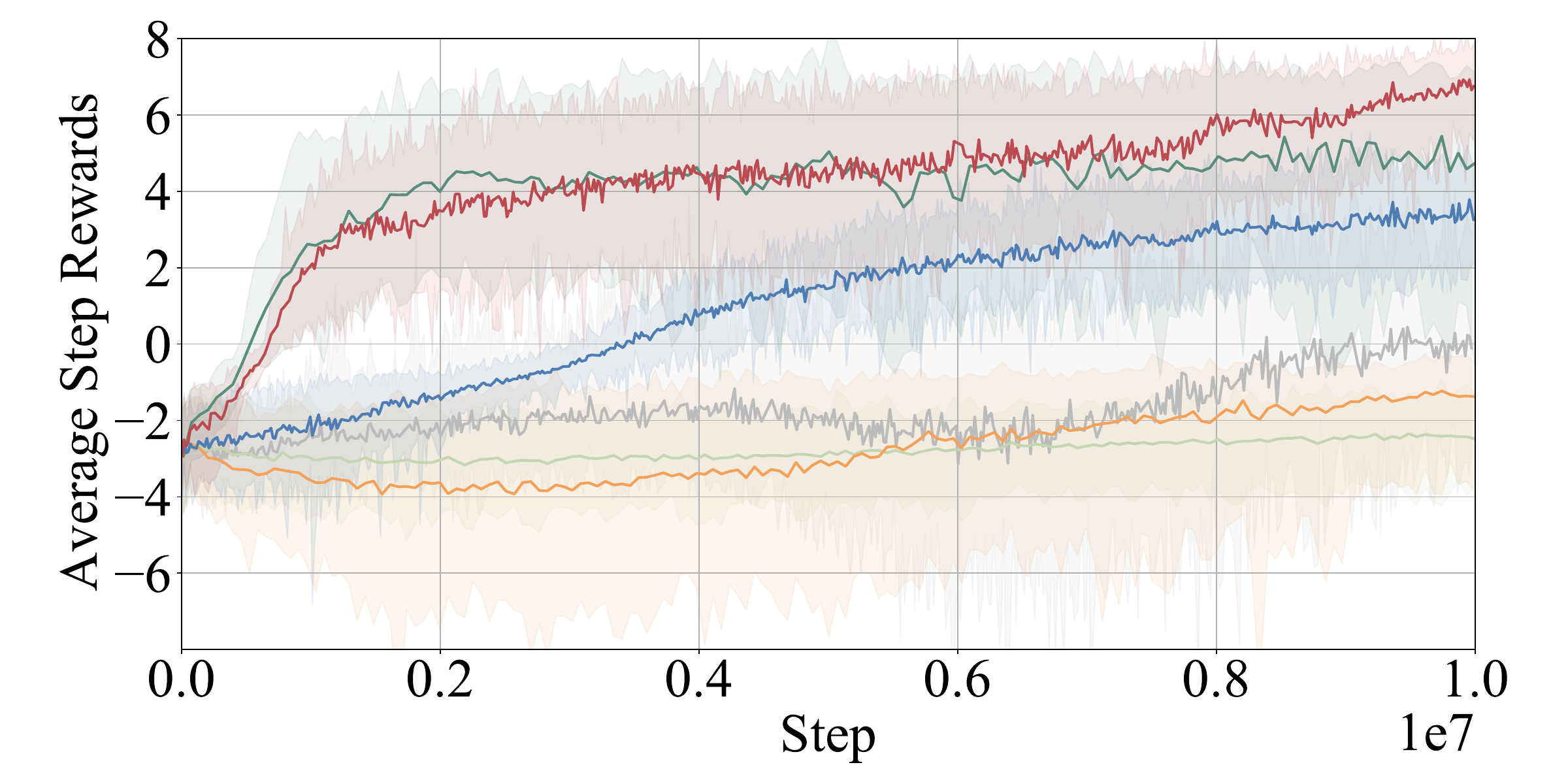}
		\caption{$n=6$}
		\label{fig:sub2}
	\end{subfigure}
	\begin{subfigure}[b]{0.30\textwidth}
		\includegraphics[width=\textwidth]{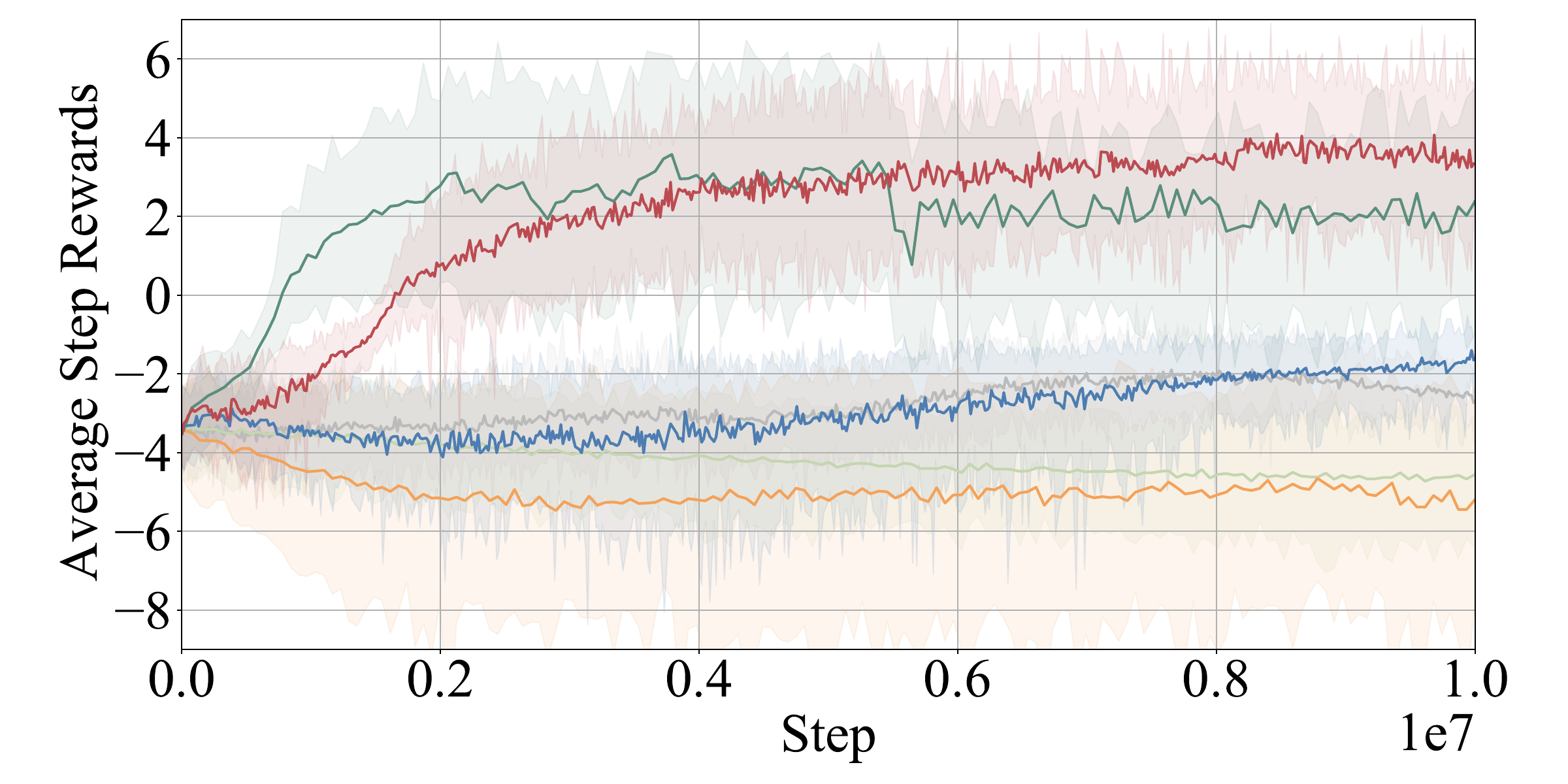}
		\caption{$n=9$}
		\label{fig:sub2}
	\end{subfigure}
	\begin{subfigure}[b]{0.38\textwidth}
		\includegraphics[width=\textwidth]{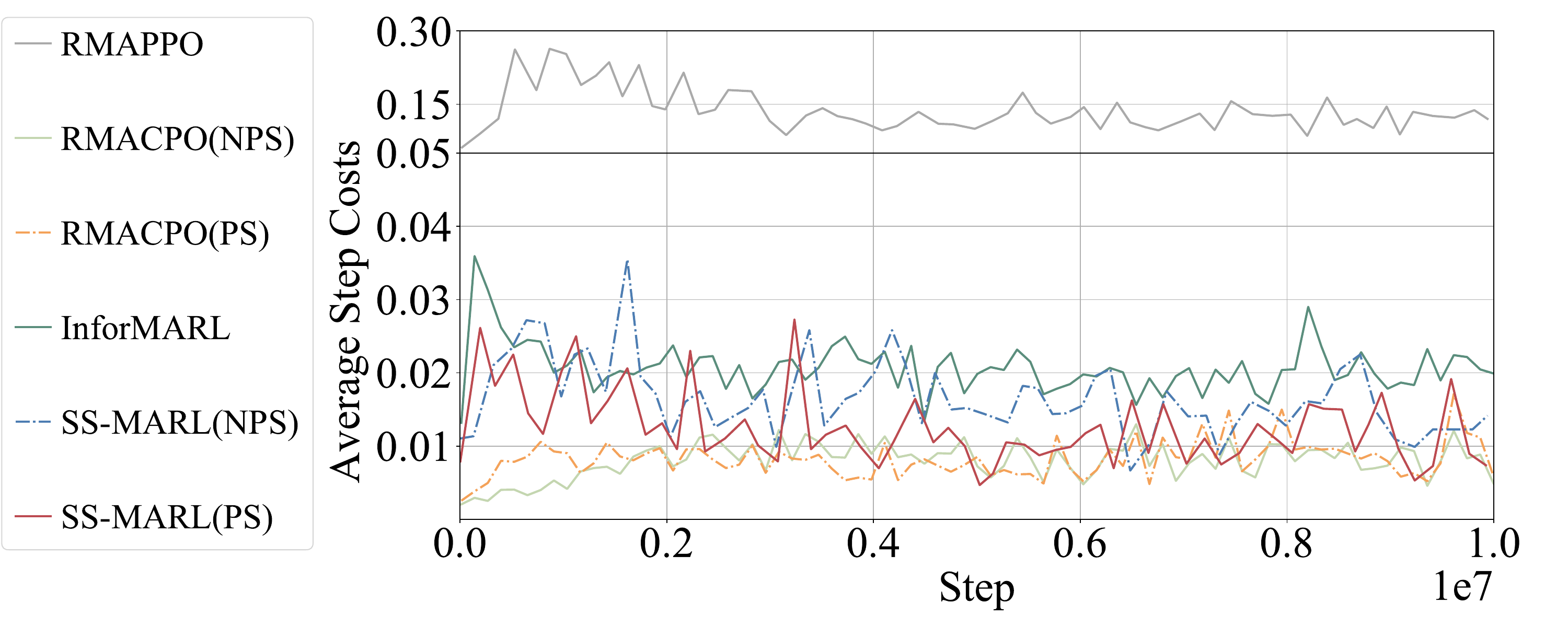}
		\caption{$n=3$}
		\label{fig:sub1}
	\end{subfigure}
	\begin{subfigure}[b]{0.30\textwidth}
		\includegraphics[width=\textwidth]{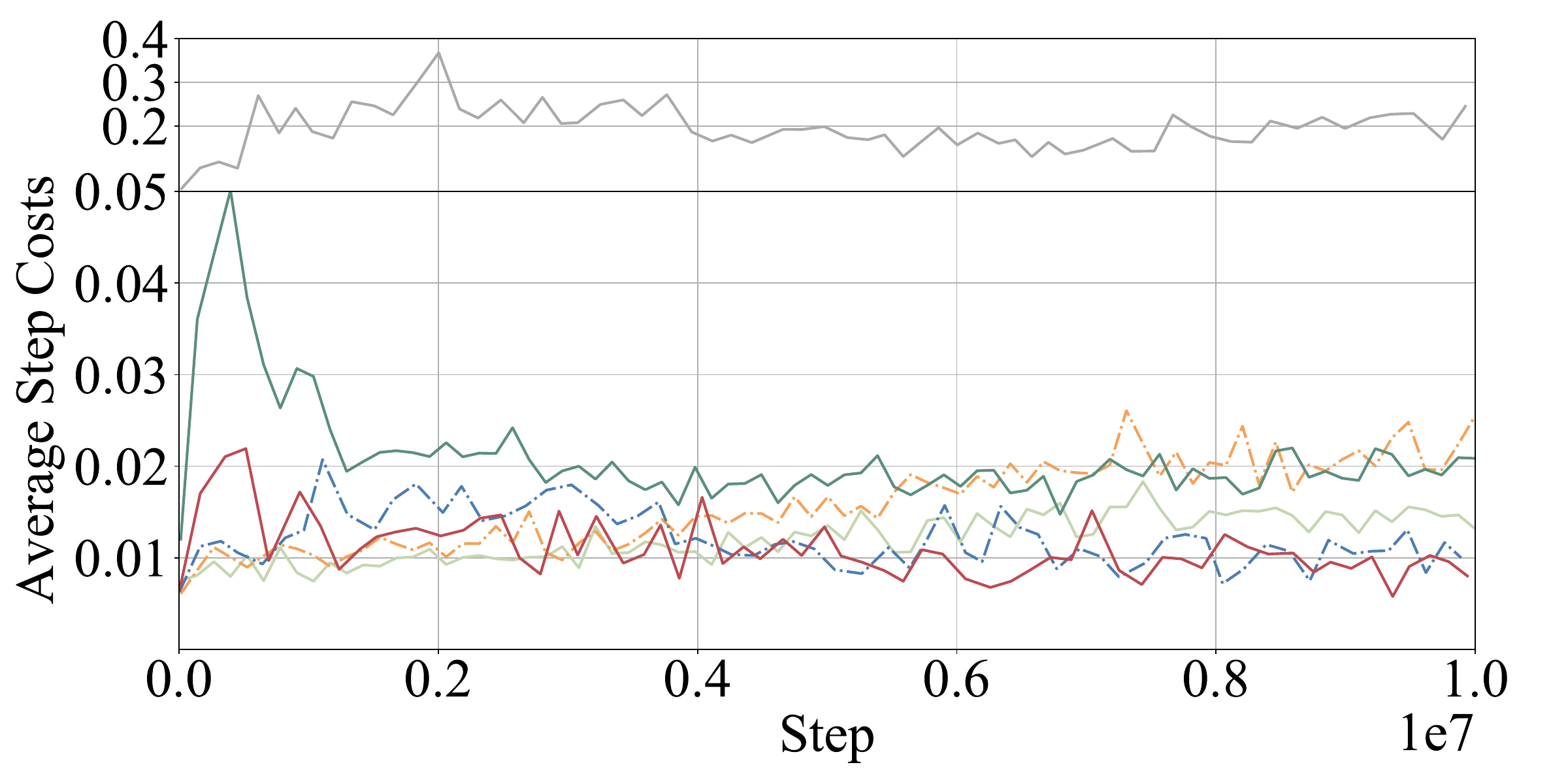}
		\caption{$n=6$}
		\label{fig:sub2}
	\end{subfigure}
	\begin{subfigure}[b]{0.30\textwidth}
		\includegraphics[width=\textwidth]{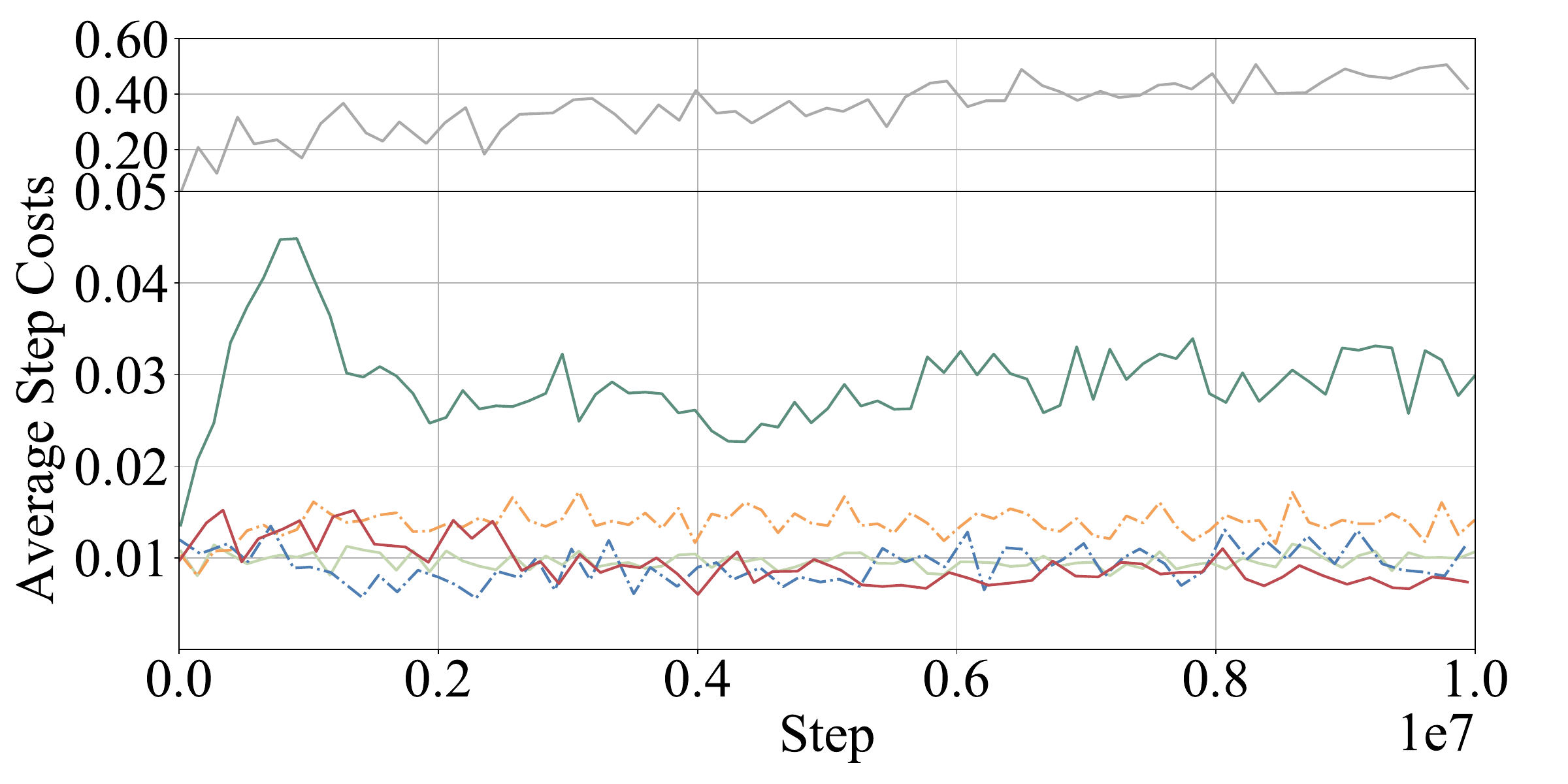}
		\caption{$n=9$}
		\label{fig:sub2}
	\end{subfigure}
	
	\caption{Comparison of the training performance of SS-MARL with baselines. (a)(b)(c) are average rewards per step per agent during the training phase, (d)(e)(f) are average costs per step per agent during the training phase.}
	\label{fig:compare}
\end{figure*}
We conduct simulation experiments in the MPE, and we have modified it to facilitate safe MARL algorithms. We selected the cooperative navigation task to validate the performance of SS-MARL, where each agent is required to reach its own goal while avoiding collisions with other entities in the environment. The number of collisions among agents is modeled as a cost, with fewer collisions indicating higher safety level. Additionally, we assume that the corresponding $c_i$ in Equation \ref{equ1} for each agent is the same value $c$, i.e., $c_i = c, \forall i \in \bm{N}$. Due to page limitations, experiments on other cooperative tasks and the hardware implementation are shown in the appendix.
\par
The subsequent experiments and analyses address three fundamental questions: 
\begin{enumerate}
	\item How does SS-MARL achieve a balance between safety and optimality?
	\item To what extent does SS-MARL enhance performance compared to other MARL algorithms?
	\item How scalable is SS-MARL in MAS?
\end{enumerate}

\subsection{SS-MARL Safety Analysis}
As specified in Equation \ref{equ1}, the cost surrogate return is constrained by an upper bound parameter, denoted as $c$. This parameter plays an important role in balancing the safety and optimality of both the training policy and the final policy trained by SS-MARL.
Accordingly, we designed our experiment as follows: within a fixed scenario featuring two agents and two obstacles. We then conduct training under two representative value of $c$ and analyze the results. \par
\begin{figure}[H]
	\centering
	\begin{subfigure}[b]{0.15\textwidth}
		\includegraphics[width=\textwidth]{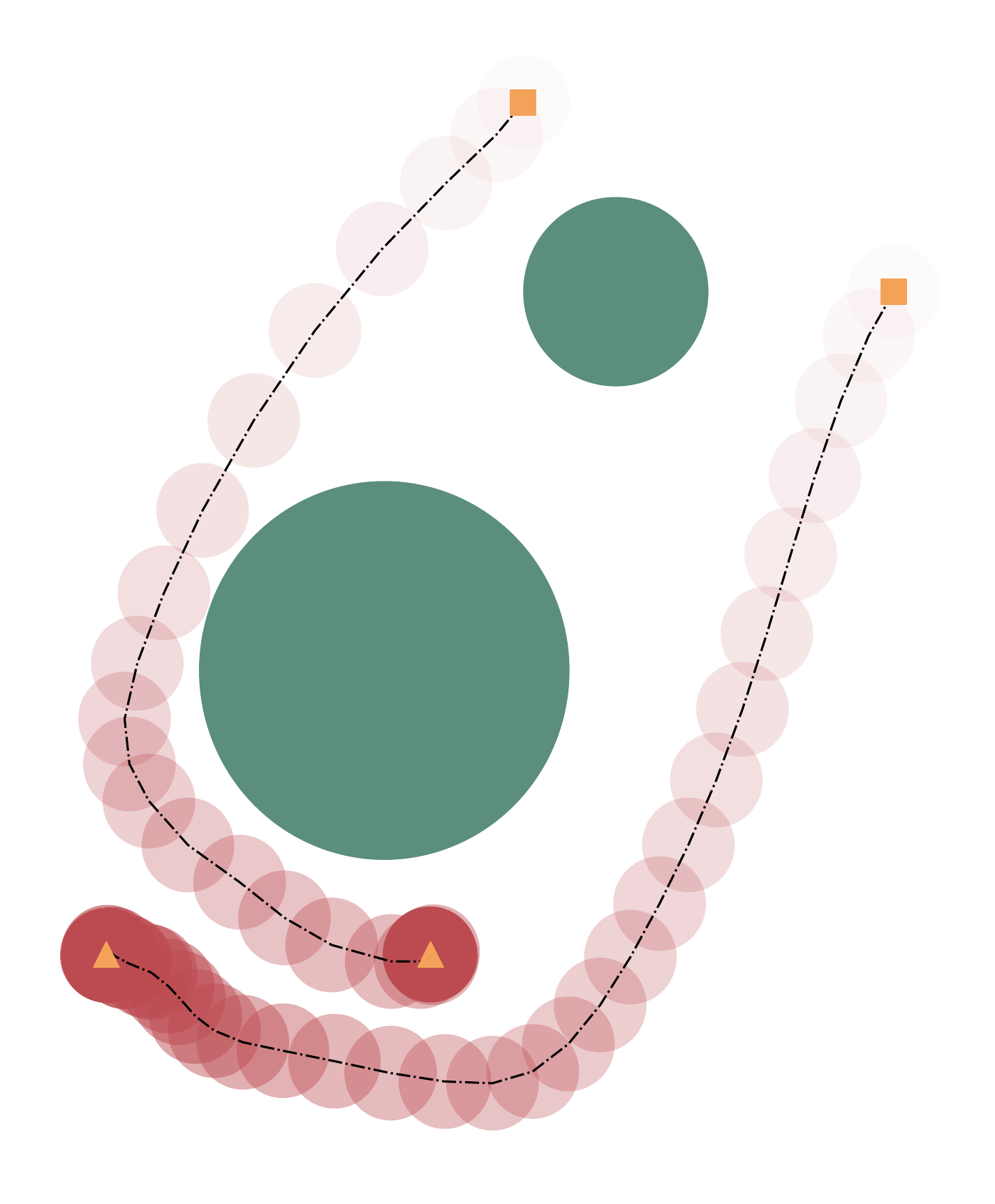}
		\caption{$c=1$}
		\label{fig:safety_sub1}
	\end{subfigure}
	\begin{subfigure}[b]{0.215\textwidth}
		\includegraphics[width=\textwidth]{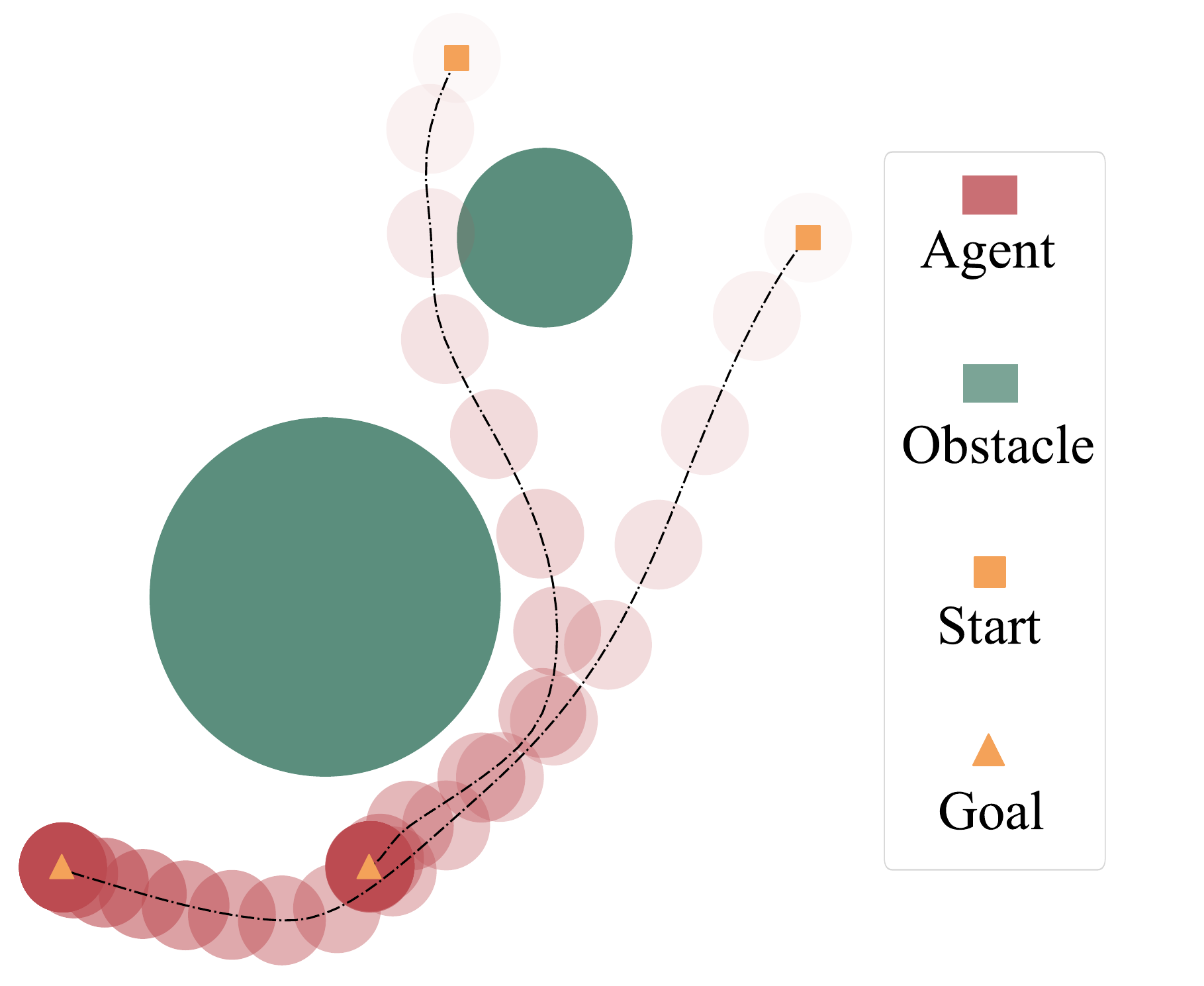}
		\caption{$c=6$}
		\label{fig:safety_sub2}
	\end{subfigure}
	\begin{subfigure}[b]{0.47\textwidth}
		\includegraphics[width=\textwidth]{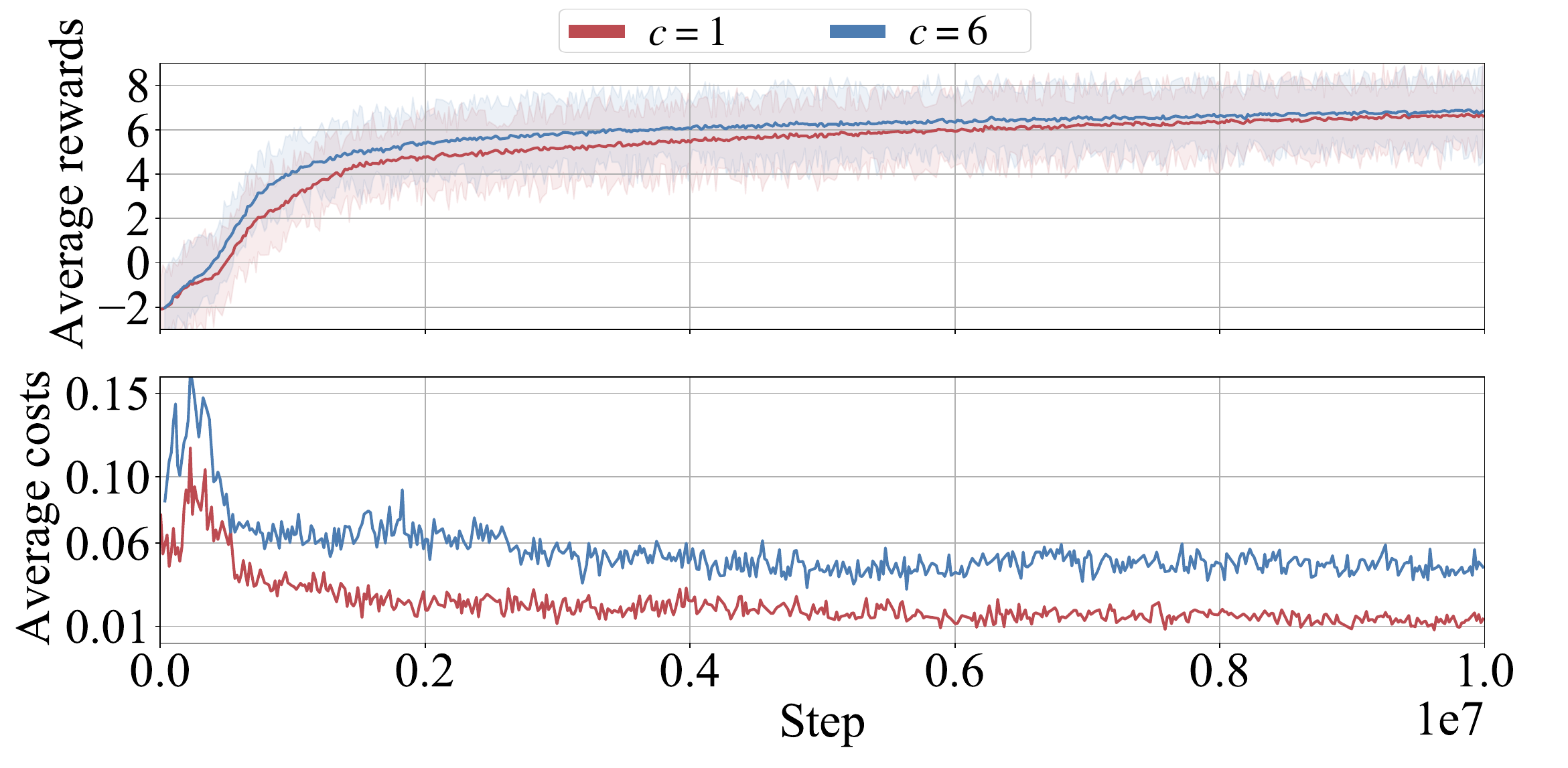}
		\caption{}
		\label{fig:safety_sub3}
	\end{subfigure}
	\caption{Scenarios with (a) $c=1$ and (b) $c=6$ and (c) average rewards per step per agent and average costs per step per agent during the training phase.}
	\label{fig:safety}
\end{figure}

From Figure \ref{fig:safety_sub3}, it can be observed that the average costs per step per agent during the training phase gradually converges around their expected upper bounds, with an episode length of 100 steps. Figures \ref{fig:safety_sub1} and \ref{fig:safety_sub2} illustrate the trajectories of final policies trained with $c=1$ and $c=6$, respectively. It is evident that the trajectory of $c=1$ is noticeably longer but safer than that of $c=6$. This is because a higher upper bound allows agents a larger exploration space, particularly regarding unsafe states. With $c=1$, the agent could not even tolerate a single collision during the training phase, leading to a trajectory that is longer but less likely to result in collisions. Additionally, the reward curves in Figure \ref{fig:safety_sub3} validate this analysis. The scenario with a smaller upper bound shows a slower convergence in its reward curve, indicating a more conservative but safer exploration process. \par
This experiment explains how SS-MARL balances safety and optimality. SS-MARL does not simply output the safest or the most optimal policy, it can achieve a balance between them by tuning adjustable parameters. Depending on the varying safety requirements of tasks within MAS, SS-MARL can be configured with different upper bounds for cost constraints to accomplish different tasks.

\subsection{Comparative Experiments}
\begin{table*}[htbp]
	\centering
	\begin{tabular}{@{}cccccccccc@{}}
		\toprule
		\multirow{2}{*}{Algorithm} & \multicolumn{3}{c}{$n=3$} & \multicolumn{3}{c}{$n=6$} & \multicolumn{3}{c}{$n=9$} \\ \cmidrule(lr){2-4} \cmidrule(lr){5-7} \cmidrule(l){8-10}
		& Reward ↑ & Cost ↓ & S/\% ↑ & Reward ↑ & Cost ↓ & S/\% ↑ & Reward ↑ & Cost ↓ & S/\% ↑ \\ \midrule
		\textbf{SS-MARL(PS)} & \textbf{2605.64} & \textbf{2.06} & \textbf{100} & \textbf{4891.89} & \textbf{4.11} & \textbf{93} & \textbf{3621.58} & \textbf{7.00} & \textbf{3} \\
		SS-MARL(NPS) & 2469.47 & 3.90 & \textbf{100} & 3284.74 & 6.49 & 38 & -1471.68 & 8.50 & 0 \\
		InforMARL & 2322.77 & 3.93 & 98 & 3476.85 & 11.65 & 52 & -3160.61 & 14.03 & 0 \\
		RMACPO(PS) & 1481.70 & 7.72 & 87 & -588.07 & 23.52 & 0 & -3831.75 & 38.20 & 0 \\
		RMACPO(NPS) & 576.97 & 8.04 & 10 & -1131.08 & 24.98 & 0 & -2929.34 & 20.03 & 0 \\
		RMAPPO & 1243.81 & 10.42 & 96 & 603.29 & 13.87 & 6 & -1689.04 & 38.06 & 0 \\ \bottomrule
	\end{tabular}
	\caption{Comparison of the testing performance of SS-MARL with baselines. Repeat the testing 100 times and take the mean. The arrival of all agents at their goals is a success. S: success rate, ↑: larger is better, ↓: smaller is better.}
	\label{Table:final_policy}
\end{table*}
The comparative experiment will be conducted in square-shaped scenarios with $n$ agents and $n$ obstacles, with a world size of $4\sqrt{n/3}$. Additionally, initial positions and goals of the agents and the positions of the obstacles in these scenarios are all randomly generated in a way that avoids conflicts. We selected $n=3, 6, 9$ to represent different levels of environmental complexity for conducting the experiments.
\par
We select these algorithms for comparative experiments: RMAPPO\cite{MAPPO}, RMACPO\cite{MACPO}, and InforMARL\cite{INFORMARL}, where 'R' stands for the RNN-based versions of the algorithms. RMACPO and RMAPPO are algorithms with fixed-size input, both of which have access to global information. In contrast, SS-MARL and InforMARL both have an identical local observation and communication radius of 1.\par
In terms of the reward function, both SS-MARL and RMACPO have their reward functions set as \( r = r_{\text{dist}} + r_{\text{goal}} \). Here \( r_{\text{dist}} \) represents the reward that guides the agent towards the goal, typically the negative of the distance to the goal. \( r_{\text{goal}} \) is a positive reward obtained when the agent reaches the goal. RMAPPO and InforMARL incorporate reward shaping to encourage obstacle avoidance by introducing an additional term, \( r_{\text{collision}} \), which represents a penalty for collisions. Consequently, their reward function is defined as \( r = r_{\text{dist}} + r_{\text{goal}} + r_{\text{collision}} \). Regarding safety parameters, SS-MARL and RMACPO set the upper bounds for their cost constraints, denoted as $c$, to 1. This means that the expected cost for an agent within one episode should not exceed 1. RMAPPO and InforMARL ensure safety by setting \( r_{\text{collision}} = -r_{\text{goal}}/2 \). It is worth noting that the setting of \( r_{\text{collision}} \) here is a result of balancing safety and optimality. Based on our experiments, further increasing the penalty does not enhance safety but instead leads to the rewards failing to converge. Additionally, we establish control groups for SS-MARL and RMACPO with policy parameter sharing, labeled as 'PS'. This indicates that each agent shares the same actor network parameter. Conversely, 'NPS' represents that each agent possesses its own actor network. \par
From Figure \ref{fig:compare}, it is evident that in all three scenarios, the reward curves of SS-MARL(PS) converge much faster than those of SS-MARL(NPS), demonstrating that policy sharing greatly enhances sampling efficiency in cooperative tasks involving homogeneous agents.
InforMARL also employs policy sharing, and its reward curve converges slightly faster than that of SS-MARL(PS). This is because SS-MARL incorporates some TRPO recovery steps when dealing with infeasible policies. These recovery steps only reduce the expected cost value, not aiming to increase the reward value. Consequently, SS-MARL requires more steps to achieve the same level of reward value. However, these recovery steps are crucial for ensuring safety. This importance is highlighted by the significantly lower cost values observed in SS-MARL compared to InforMARL during training. Particularly in the more complex scenario with $n=9$, the difference is even more pronounced.
In terms of cost value metrics, SS-MARL performs slightly better than RMACPO. However, there is a significant difference in the convergence of rewards. SS-MARL, which utilizes GNN for agent communication, demonstrates a marked advantage in this regard. In contrast, RMAPPO performs even worse because the reward shaping approach fails to provide safety guarantees. Furthermore, in complex scenarios, RMAPPO cannot even achieve convergence in rewards.\par
Regarding the final policy, Table \ref{Table:final_policy} shows that SS-MARL(PS) has a significant advantage in terms of reward value, cost value, and success rate. Notably, when $n=9$, other algorithms fail to achieve positive rewards, indicating that most agents do not reach their goals. On the contrary, SS-MARL not only obtains positive rewards but also satisfies the cost constraint.

\subsection{Scalability Experiments}
To test the scalability of SS-MARL, we selected the models that were trained with $n=3$ and \textbf{zero-shot transfer} to scenarios with larger $n$, and compared the results with those of two state-of-the-art methods: InforMARL \cite{INFORMARL} and TEM \cite{TEM[IJCAI]}. As the world size grows with larger  $n$, the time required for an agent to reach the goal also increases. Since we will test on scenarios with a maximum $n=96$, to prevent the episode from being too short and causing agents to fail in completing the task, we extended the episode length to 200 steps.
\begin{figure}[H]
	\centering
	\includegraphics[width=0.49\textwidth]{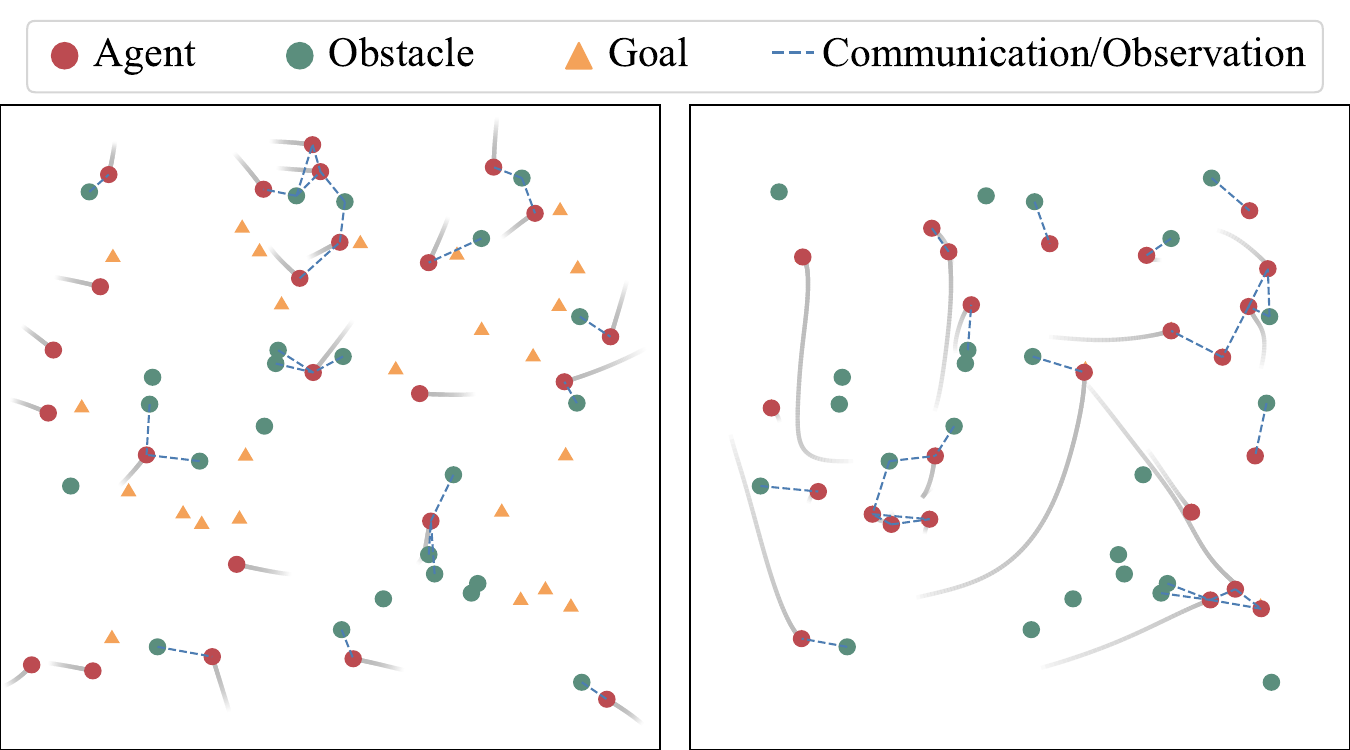}
	\caption{Zero-shot transfer to $n=24$ using a model trained on scenarios with $n=3$, when the test episode (left) begins and (right) ends}
	\label{fig:24agents}
\end{figure}
During the testing phase, the graph that represents the agent's communication and observation undergoes dynamic changes. Concurrently, the attention weights associated with the communication and observation edges dynamically adjust. This adjustment helps to balance the motion towards the goal with the avoidance of local unsafe states. Figure \ref{fig:24agents} illustrates the testing scenario with $n=24$, using a model trained on scenarios with only $n=3$. The gray curves with gradient color represent the trajectories of the agents. These trajectories indicate that the agents not only complete global navigation tasks but also possess great local obstacle avoidance capabilities.
\begin{figure}[H]
	\centering
	\includegraphics[width=0.49\textwidth]{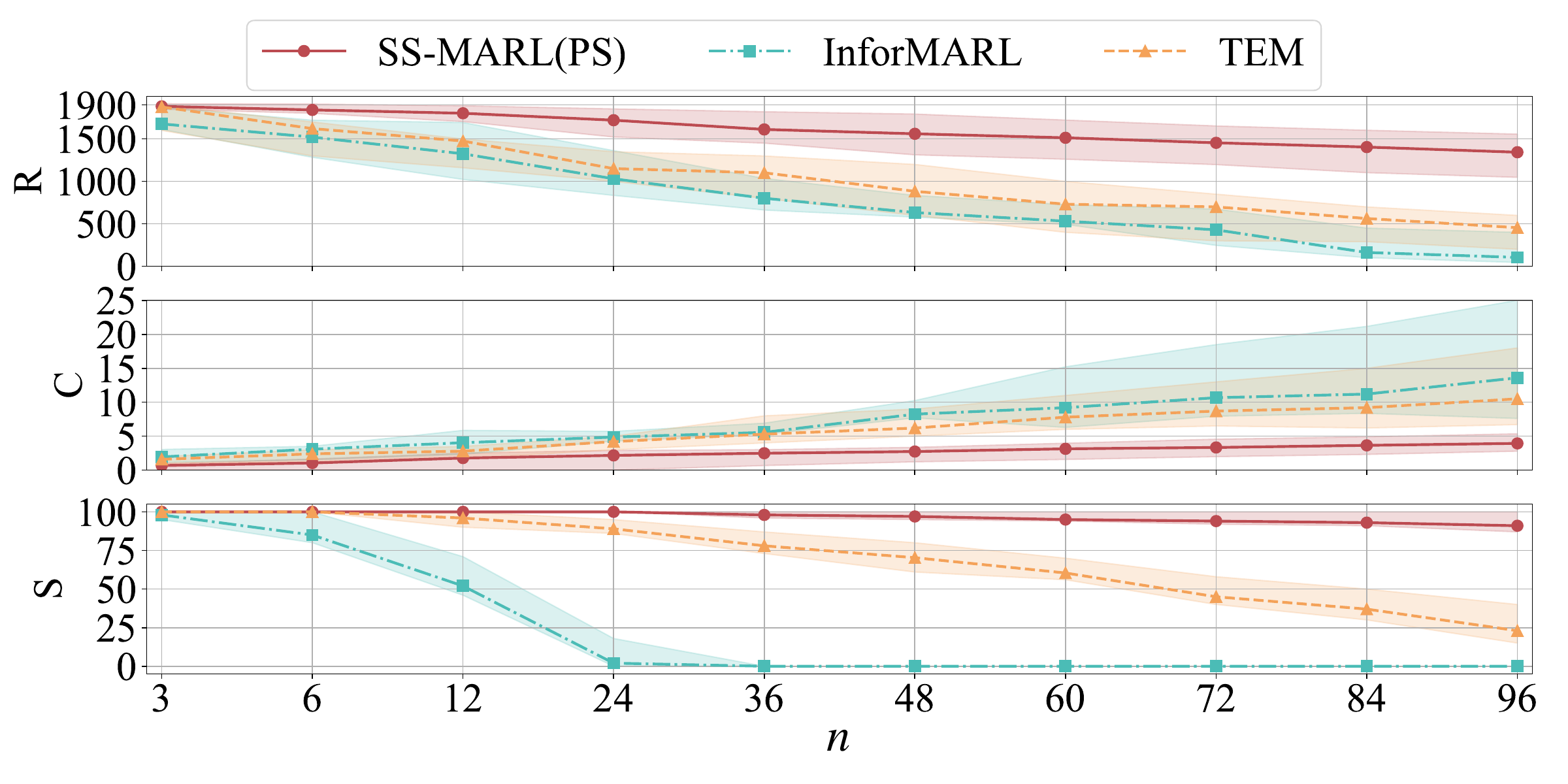}
	\caption{Comparison of the zero-shot transfer performance of SS-MARL and other baselines. Repeat the testing 100 times on every $n$ and take the mean. The model is trained on scenarios with $n=3$. R: average episode rewards per agent, C: average episode costs per agent, S: success rate.}
	\label{fig:exp3}
\end{figure}
From Figure \ref{fig:exp3}, it is evident that SS-MARL demonstrates strong scalability, outperforming the other two methods in all metrics. Notably, the SS-MARL model trained with only 3 agents can achieve a success rate exceeding 90\% and maintain collision times per agent below 5 when scaled up to 96 agents. These experimental results demonstrate the strong scalability of SS-MARL.
\section{Conclusion}
This paper introduces SS-MARL, a novel approach to MARL that emphasizes safety and scalability. SS-MARL is designed to handle complex cooperative tasks in MAS where agents must comply with various local and global safety constraints. We introduced constrained joint policy optimization to achieve the safety of both the training policy and the final policy. Additionally, GNNs endows SS-MARL with strong scalability. Experimental results demonstrate that SS-MARL achieves the best balance between optimality and safety compared to the baselines. Compared to the latest methods, it also shows great scalability in scenarios with a large number of agents. By leveraging SS-MARL, the MAS trained through MARL can become safer and more scalable, enhancing the potential for MARL applications in real-world tasks.

\bibliographystyle{named}
\bibliography{ijcai25}

\end{document}


\maketitle
	\section*{Overview}
	To provide comprehensive details of our algorithm and experiments, we further present more theoretical and experimental details of Scalable Safe Multi-Agent Reinforcement Learning (SS-MARL) in the supplementary material. The additional content is presented from the following perspectives.
	\begin{itemize}
		\item \textbf{Theoretical Details}
		\begin{itemize}
			\item {Monotone Improvement of the Constrained Joint Policy Optimization}
			\item {Practical Implementation}
		\end{itemize}
		\item \textbf{Experimental Details}
		\begin{itemize}
			\item {Hyperparameters}
			\item {Safe MPE: an Evironment for Safe MARL algorithms}
			\item {Supplementary Simulation Experiments}
			\item {Supplementary Hardware Experiments}
		\end{itemize}
	\end{itemize}
	
	\section{Theoretical Details}
	Here, we primarily undertake the proof of the Joint Policy convergence, that is, demonstrating that the Joint Policy can monotonically improve while satisfying the constraints. Additionally, we point out where SS-MARL makes approximations and discuss the underlying theory that SS-MARL utilizes.
	\subsection{Monotone Improvement of the Constrained Joint Policy Optimization}
	The original optimization problem is shown in paper as below:
	\begin{align}
		\pi^{k+1}_{i_h} &= \arg\max_{\pi} L_{\bm{\pi}^{k}}^{i_{1: h}}\left(\bm{\pi}^{k+1}_{i_{1: h-1}}, \pi\right)-\nu D_{\mathrm{KL}}^{\mathrm{max}}(\pi, \pi_{i_h}^k), \nonumber \\
		&\text{s.t.   } D_{\text{KL}}^{\text{max}}\left(\pi^{k}_{i_h}, \pi\right) \leq \delta_{i_h},\nonumber \\
		&J_{\text{c},i_{h}}^{j}\left(\bm{\pi}^{k}_{i_{1:h-1}}\cup \pi\right)\leq c^{j}_{i_{h}} , \forall j = 1, \ldots, m_{i_{h}}.
		\label{opt1}
	\end{align}
	So, subsequent theoretical derivations were made to show that the policy updated according to this optimization problem is monotonically improved.
	\begin{lemma}\label{lemma:MAAD}
		Multi-Agent Advantage Decomposition \cite{MAAD}. For any state $S$ , subsets $i_{1:h}\subseteq\bm{N}$ and joint action $\bm{A}_{i_{1:h}}$, the following equation holds:
	\begin{align}
		\mathcal{A}_{\bm{\pi}}^{i_{1: h}}\left(S, \bm{A}_{i_{1: h}}\right) =  \sum_{j=1}^{h} \mathcal{A}_{\bm{\pi}}^{i_{j}}\left(S, \bm{A}_{i_{1: j-1}}, A_{i_{j}}\right).
	\end{align}
	\proof As the definition of Multi-Agent Advantage Function, it can be expanded as below.
	\begin{align}
		\mathcal{A}_{\bm{\pi}}^{i_{1: h}}\left(S, \bm{A}_{i_{1: h}}\right) &=  Q_{\bm{\pi}}^{i_{1:h}}\left(S, \bm{A}_{i_{1: h}} \right) - V_{\bm{\pi}}(S) \nonumber \\
		& = \sum_{j=1}^{h}[Q_{\bm{\pi}}^{i_{1:j}}\left(S, \bm{A}_{i_{1: j}} \right)-Q_{\bm{\pi}}^{i_{1:j-1}}\left(S, \bm{A}_{i_{1: j-1}} \right)] \nonumber \\
		& =
		\sum_{j=1}^{h}\mathcal{A}_{\bm{\pi}}^{i_{j}}\left(S, \bm{A}_{i_{1: j-1}}, A_{i_{j}}\right).
	\end{align}
		
	\end{lemma}

	According to Lemma \ref{lemma:MAAD}, we can expand the definition of joint surrogate return as follow\footnote{Note: here $j$ is used to represent the $j$th cost constraints of agent $i$}:
	\begin{align}
		L_{\bm{\pi}}^{i}(\bar{\pi}_i)\triangleq  \mathbb{E}_{S\sim\rho_0,A_{i}\sim\bar{\pi}_{i}} \left[\mathcal{A}_{\bm{\pi}}^{i} \left(S, A_{i}\right)\right].
	\end{align}
	\begin{align}\label{equ:L_cost}
		L_{\text{c},\bm{\pi}}^{i,j}(\bar{\pi}_i)\triangleq  \mathbb{E}_{S\sim\rho_0,A_{i}\sim\bar{\pi}_{i}} \left[\mathcal{A}_{\text{c},\bm{\pi}}^{i,j} \left(S, A_{i}\right)\right], \forall j=1,...,m^i.
	\end{align}
	
	Following we can give another lemma.
	
	\begin{lemma}\label{lemma:2}
		Costs Bounds \cite{MACPO}. Let $\bm{\pi}$ and $\bar{\bm{\pi}}$ be two different joint policies, the following inequality holds
		\begin{align}\label{equ:5}
			J_{\mathrm{c},i}^j(\bar{\bm{\pi}})\leq J_{\mathrm{c},i}^j(\bm{\pi}) + L_{\mathrm{c},\bm{\pi}}^{i,j}(\bar{\pi}_i)+\nu_i^j\sum_{h=1}^{n}D_{\mathrm{KL}}^{\mathrm{max}}(\pi_h,\bar{\pi}_h),\nonumber\\
			\mathrm{where}\quad \nu_i^j = \frac{4\gamma_{\mathrm{c}}\max_{S,A_i}|\mathcal{A}_{\mathrm{c},\bm{\pi}}^{i,j}\left(S, A_{i}\right)|}{(1-\gamma_{\mathrm{c}})^2}.
		\end{align}
	\proof Apply TRPO bound \cite{TRPO} in joint policies, and use Equation \ref{equ:L_cost}. We can conclude that:
	\begin{align}
		J_{\mathrm{c},i}^j(\bar{\bm{\pi}})\leq J_{\mathrm{c},i}^j(\bm{\pi}) + L_{\mathrm{c},\bm{\pi}}^{i,j}(\bar{\pi}_i)+\frac{4\alpha^2\gamma_{\mathrm{c}}\max_{S,A_i}|\mathcal{A}_{\mathrm{c},\bm{\pi}}^{i,j}\left(S, A_{i}\right)|}{(1-\gamma_{\mathrm{c}})^2}.
	\end{align}
	The key lies in $\alpha^2$, and  $\alpha = D_{\mathrm{TV}}^{\mathrm{max}}(\bm{\pi}_i,\bar{\bm{\pi}}_i)$. In TRPO, there is an important inequality $D_{\mathrm{TV}}(p,q)^2\leq D_{\mathrm{KL}}(p,q)$. So, we can get:
	\begin{align}\label{equ:7}
		J_{\mathrm{c},i}^j(\bar{\bm{\pi}})\leq J_{\mathrm{c},i}^j(\bm{\pi}) + L_{\mathrm{c},\bm{\pi}}^{i,j}(\bar{\pi}_i)+\nu_i^jD_{\mathrm{KL}}^{\mathrm{max}}(\bm{\pi},\bar{\bm{\pi}}).
	\end{align}
	Notice that:
	
	\begin{align}\label{equ:8}
		D_{\mathrm{KL}}^{\mathrm{max}}(\bm{\pi},\bar{\bm{\pi}}) = \max_S\left(\sum_{l=1}^{n}D_{\mathrm{KL}}(\pi_l(\cdot|S),\bar{\pi}_l(\cdot|S))\right)\leq\sum_{l=1}^{n}\max_SD_{\mathrm{KL}}(\pi_l(\cdot|S),\bar{\pi}_l(\cdot|S))=\sum_{h=1}^{n}D_{\mathrm{KL}}^{\mathrm{max}}(\pi_h,\bar{\pi}_h).
	\end{align}
	By combining inequations \ref{equ:7} and \ref{equ:8} we can finally get inequation \ref{equ:5} proved.
	
	\end{lemma}
	
	\begin{theorem}
		Monotone Improvement of the Constrained Joint Policy Optimization. If $\{\bm{\pi}^0, \bm{\pi}^1,...,\bm{\pi}^k,...\}$ is a sequence of joint policies obtained from the optimization problem \ref{opt1}. Then $J(\bm{\pi}^{k+1})\ge J(\bm{\pi}^{k})$, as well as it satisfies the cost constraints, $J_{\mathrm{c},i}^j(\bm{\pi}^{k})\le c_i^j, \forall k\in\mathbb{N}, i\in \bm{N} \text{ and } j\in\{ 1,...,m^i\}$.
		\proof According to Lemma \ref{lemma:2}, $J_{\mathrm{c},i}^j(\bm{\pi}^{k+1})$ is limited by a upper bound. If we consider $\bm{\pi}^{k+1}$ to be a joint policy updated from $\bm{\pi}^{k}$. So the upper bound is dominated by the KL distance between $\bm{\pi}^{k+1}$ and $\bm{\pi}^{k}$, we can actually control the upper bound by limit the KL distance. The specific form of this limit is given in \cite{MACPO}. By choosing a suitable KL distance limit, the cost constraint can be satisfied. From \cite{TRPO}, we have
		\begin{align}
			J(\bm{\pi}^{k+1})\ge J(\bm{\pi}^{k}) +  \mathbb{E}_{S\sim\rho_0,\bm{A}\sim\bm{\pi}^{k+1}} \left[\mathcal{A}_{\bm{\pi}^k} \left(S, \bm{A}\right)\right] - \nu D_{\mathrm{KL}}^{\mathrm{max}}(\bm{\pi}^k, \bm{\pi}^{k+1}).
		\end{align} 
		It is lower-bounded by Equation \ref{equ:8} :
		\begin{align}
			J(\bm{\pi}^{k+1})\ge J(\bm{\pi}^{k}) +  \mathbb{E}_{S\sim\rho_0,\bm{A}\sim\bm{\pi}^{k+1}} \left[\mathcal{A}_{\bm{\pi}^k} \left(S, \bm{A}\right)\right] - \sum_{h=1}^{n}\nu D_{\mathrm{KL}}^{\mathrm{max}}(\pi_{i_h}^k, \pi_{i_h}^{k+1}).
		\end{align} 
		According to Lemma \ref{lemma:MAAD}, it can be expanded as:
		\begin{align}\label{equ:11}
			J(\bm{\pi}^{k+1})&\ge J(\bm{\pi}^{k}) + \sum_{h=1}^{n} \mathbb{E}_{S\sim\rho_0,\bm{A}_{i_{1:h}}\sim\bm{\pi}^{k+1}_{i_{1:h}}} \left[\mathcal{A}_{\bm{\pi}^k}^{i_h} \left(S, \bm{A}_{i_{1:h-1}}, A_{i_h}\right)\right] - \sum_{h=1}^{n}\nu D_{\mathrm{KL}}^{\mathrm{max}}(\pi_{i_h}^k, \pi_{i_h}^{k+1}) \nonumber \\
			&= J(\bm{\pi}^{k}) + \sum_{h=1}^{n}\left(L_{\bm{\pi}^k}^{i_{1:h}}(\bm{\pi}_{i_{1:{h-1}}}^{k+1}, \pi_{i_h}^{k+1})-\nu D_{\mathrm{KL}}^{\mathrm{max}}(\pi_{i_h}^k, \pi_{i_h}^{k+1}) \right).
		\end{align} 
		And because $\pi_{i_h}^{k+1}$ is obtained by the optimization problem \ref{opt1} which is the argmax of the objective, we can obtain:
		\begin{align} \label{equ:12}
			L_{\bm{\pi}^k}^{i_{1:h}}(\bm{\pi}_{i_{1:{h-1}}}^{k+1}, \pi_{i_h}^{k+1})-\nu D_{\mathrm{KL}}^{\mathrm{max}}(\pi_{i_h}^k, \pi_{i_h}^{k+1}) \ge L_{\bm{\pi}^k}^{i_{1:h}}(\bm{\pi}_{i_{1:{h-1}}}^{k+1}, \pi_{i_h}^{k})-\nu D_{\mathrm{KL}}^{\mathrm{max}}(\pi_{i_h}^k, \pi_{i_h}^{k}).
		\end{align}
		According to Equation \ref{equ:11} and \ref{equ:12}, we can obtain:
		\begin{align}
			J(\bm{\pi}^{k+1})\ge J(\bm{\pi}^{k}) + \sum_{h=1}^{n}\left(L_{\bm{\pi}^k}^{i_{1:h}}(\bm{\pi}_{i_{1:{h-1}}}^{k+1}, \pi_{i_h}^k)-\nu D_{\mathrm{KL}}^{\mathrm{max}}(\pi_{i_h}^k, \pi_{i_h}^k) \right)
		\end{align}
		As the definition of joint surrogate return and KL distance, we can finally get:
		\begin{align}
			J(\bm{\pi}^{k+1})\ge J(\bm{\pi}^{k}) + \sum_{h=1}^{n}0 =J(\bm{\pi}^{k})
		\end{align}
		So, the monotone improvement of the Constrained Joint Policy Optimization is proved.
	\end{theorem}
	\subsection{Practical Implementation}
	Although the theoretical monotone improvement is proved, there are approximations in practical implementation. In practical implementation, we parameterize the joint policy \( \bm{\pi}^k = \prod_{i=1}^{n}\pi_{\theta_i^k} \) after $k$ updates. Thus, the optimization problem \ref{opt1} can be simplified to the following form:
	\begin{align}
		\theta^{k+1}_{i_h} &= \arg\max_{\theta} \mathbb{E}_{S \sim \rho_0, \bm{A}_{i_{-h}} \sim \bm{\pi}^{k}_{i_{-h}}, A_{i_{h}} \sim \pi_{\theta}}\left[\mathcal{A}_{\bm{\pi}^{k}}^{i_{h}}\left(S, \bm{A}_{i_{-h}}, A_{i_{h}}\right)\right] \nonumber \\
		& \text{s.t. } J_{\text{c},i_h}^{j}\left(\bm{\pi}^{k}\right) + \mathbb{E}_{S \sim \rho_0, A_{i_{h}} \sim \pi_{\theta}}\left[\mathcal{A}_{\text{c},\bm{\pi}^{k}}^{i_h,j}\left(S, A_{i_{h}}\right)\right] \leq c^{j}_{i_{h}}, \nonumber \\
		&\forall j = 1, \ldots, m_{i_{h}}, \text{ and } \bar{D}_{\text{KL}}\left(\pi_{\theta^{k}_{i_h}}, \pi_{\theta}\right) \leq \delta.
		\label{opt2}
	\end{align}
	We can use Taylor expansion to further approximate this problem:
\begin{align}
	\theta^{k+1}_{i_h} &= \arg\max_{\theta}(\bm{G}_{i_h})(\theta-\theta^{k}_{i_h}) \nonumber \\
	& \text{s.t. } D_{i_h}^j + (\bm{B}_{i_h}^j)(\theta-\theta^{k}_{i_h})\le 0, \forall j = 1, \ldots, m_{i_{h}},\nonumber \\
	& \text{and } \frac{1}{2}(\theta-\theta^{k}_{i_h})^T\bm{E}_{i_h} (\theta-\theta^{k}_{i_h}) \leq \delta.
	\label{opt3}
\end{align}
	\par
	Here, $\bm{G}_{i_h}$ is the gradient of the objective of agent $i_h$, $\bm{B}_{i_h}^j$ is the gradient of $j$th cost constraint of agent $i_h$, $\bm{E}_{i_h}$ is the Hessian matrix of the average KL distance of agent $i_h$.
	Similar to \cite{TRPO,HATRPO,MACPO}, we can take a primal-dual approach to solve this linear quadratic optimization problem. \par
	However, this linear quadratic optimization problem is only an approximation, the policy $\theta^{k}_{i_h}$ still might be infeasibe. \cite{CPO} and \cite{MACPO} gives a recovery step when $m^{i}=1, \forall i\in \bm{N}$. But their algorithms do not give the solution of multiple cost constraints. We will give the solution of multiple cost constraints. With multiple cost constraints functions, the recovery step is actually a multi-objective constrained optimization problem shown as below:
	\begin{align}
		&\min_\theta (\bm{B}_{i_h}^{1})(\theta-\theta^{k}_{i_h}) \nonumber \\
		&\min_\theta (\bm{B}_{i_h}^{2})(\theta-\theta^{k}_{i_h}) \nonumber \\
		&... \nonumber \\
		&\min_\theta (\bm{B}_{i_h}^{m_{i_h}})(\theta-\theta^{k}_{i_h}) \nonumber \\
		&\text{s.t.  } \frac{1}{2}(\theta-\theta^{k}_{i_h})^T\bm{E}_{i_h} (\theta-\theta^{k}_{i_h}) \leq \delta.
	\end{align}
	
	There is no need to accurately solve the Pareto front of this multi-objective constrained optimization problem, because we only require that the constraints can be reduced. Thus, we construct a weighted penalty function to transform a multi-objective problem into a single-objective problem shown as below:
	\begin{align}
		&\min_\theta \sum_{j=1}^{m_{i_h}}\left(\beta_j\bm{B}_{i_h}^{j}\right)(\theta-\theta^{k}_{i_h}) \nonumber \\
		&\text{s.t.  } \frac{1}{2}(\theta-\theta^{k}_{i_h})^T\bm{E}_{i_h} (\theta-\theta^{k}_{i_h}) \leq \delta.
	\end{align}
	
	$\beta_j$ is an important weight, it is used to describe the importance of the cost constraints. Constraints that exceed the constraint value by a lot require more weight. In contrast, those that satisfy the constraints value do not need to be reduced and require weight to be zero. So, we can write $\beta_j$ as below:
	\begin{align}
		\beta_j&=\left\{\begin{matrix} 
			\frac{\text{exp}(P_j)}{\sum_{q\in u}\text{exp}(P_q)}  \text{ , }j\in u\\  
			0   \text{ , }j\notin u
		\end{matrix}\right. \nonumber\\
		P_j &= J_{\text{c},i_h}^{j}\left(\bm{\pi}^{k}\right) - c_i^j, \nonumber\\
		u&=\left\{j|P_j>0\right\}.
	\end{align}
	
	With the weighted penalty function, we can use a TRPO step \cite{TRPO} on weighted matrix $\sum_{j=1}^{m_{i_h}}\beta_j\bm{B}_{i_h}^{j}$ to decrease the cost constraints and recover the policy.
	
	\begin{align}
		&\mathcal{B}_{i_h} = \sum_{j=1}^{m_{i_h}}\beta_j\bm{B}_{i_h}^{j}\nonumber \\
		\theta^{k+1}_{i_h}=\theta^k_{i_h}- \alpha_j& \sqrt{\frac{2\delta}{(\mathcal{B}_{i_h}) \left(\bm{E}_{i_h}\right)^{-1} (\mathcal{B}_{i_h})^T}}\left(\bm{E}_{i_h}\right)^{-1} (\mathcal{B}_{i_h})^T
	\end{align}
	
	$\alpha_j$ is adjusted through backtracking line search. Thus, even if an infeasible solution is generated during the update process, the infeasible solution can still be recovered to a feasible solution by a recovery step. Meanwhile, it is different from the simpler assumptions in CPO\cite{CPO} and MACPO\cite{MACPO}, SS-MARL can recover the joint policy even if $m_i>1$.
	
	\section{Experimental Details}
	\subsection{Hyperparameters}
	Tables  show the hyperparameters for SS-MARL.
	
	\begin{minipage}[t]{0.48\textwidth} 
		\centering 
		\captionof{table}{Network related hyperparameters.} 
		\begin{tabular}{lc}
			\toprule			Hyperparameters             & Value      \\ \midrule
			Embedding number            & 3          \\
			Embedding dimension         & 4          \\
			Embedding layer hidden size & 16         \\
			GNN layer number            & 2          \\
			GNN edge dimension          & 4          \\
			GNN attention heads number  & 2          \\
			GNN activation              & ReLU       \\
			GNN out dimension           & 64         \\
			LSTM layer number           & 1          \\
			LSTM time length of chunks  & 10         \\
			Network initialisation      & Orthogonal \\ \bottomrule
		\end{tabular}
\end{minipage}%
\hfill 
\begin{minipage}[t]{0.48\textwidth} 
	\centering
	\captionof{table}{Training related hyperparameters.} 
		\begin{tabular}{lc}
			\toprule
			Hyperparameters              & Value      \\ \midrule
			gradient clip norm           & 10.0       \\
			gae lambda                   & 0.95       \\
			gamma                        & 0.99       \\
			cost gamma                   & 0.99       \\
			value loss                   & huber loss \\
			cost loss                    & huber loss \\
			huber delta                  & 10.0       \\
			line search fraction         & 0.5        \\
			fraction coefficient         & 0.05       \\
			mini batch number            & 2          \\
			optimizer                    & Adam       \\
			optimizer epsilon            & 1e-5       \\
			learning rate decay          & False      \\
			learning rate                & 7e-3       \\
			buffer length                & 100        \\
			training environments number & 32         \\ \bottomrule
		\end{tabular}
		\label{Table:final_policy}
	\end{minipage}
	\subsection{Safe MPE: an Evironment for Safe MARL Algorithms}
	The original Multi-Agent Particle Environment (MPE) \cite{MADDPG} is a simple multi-agent particle world with a continuous observation and discrete action space, along with some basic simulated physics. MPE can be used to evaluated MARL algorithms. But MPE itself does not carry cost constraint information which is not applicable to Safe MARL algorithms. Therefore, we add information about cost constraints in the environment on the basis of MPE, and the resulting environment is called Safe MPE. \par
	Safe MPE can be used for different kinds of Safe MARL algorithms. We have defined three types of environments to accommodate different Safe MARL algorithms shown as below.
	
	\begin{itemize}
		\item \textbf{MultiAgentEnv:}
		\begin{itemize}
			\item Base class for other environment classes.
			\item Only fixed size observations of the agent are allowed.
			\item Does not consider cost constraints.
			\item Suitable for basic MARL algorithms, such as MAPPO \cite{MAPPO}, MADDPG \cite{MADDPG}, MATD3 \cite{MATD3}, etc.
		\end{itemize}
		\item \textbf{MultiAgentConstrainEnv:}
		\begin{itemize}
			\item Only fixed size observations of the agent are allowed.
			\item Consider cost constraints.
			\item Suitable for basic Safe MARL algorithms, such as MACPO \cite{MACPO}, MAPPO-Lagrangian \cite{MACPO}, etc.
		\end{itemize}
		\item \textbf{MultiAgentGraphConstrainEnv:}
		\begin{itemize}
			\item Variable-sized observations setting in SS-MARL are allowed.
			\item Consider cost constraints.
			\item Suitable for SS-MARL.
		\end{itemize}
	\end{itemize}
	
	Users can use the Cooperative Navigation (CN) scenario described in the paper or create and use custom scenarios. As shown in this figure, the environment will visualize the agents, obstacles, goals, communication and observation. And users can test their own Safe MARL algorithms on these scenarios.
	\begin{figure}[H]
		\centering
		\includegraphics[width=0.7\textwidth]{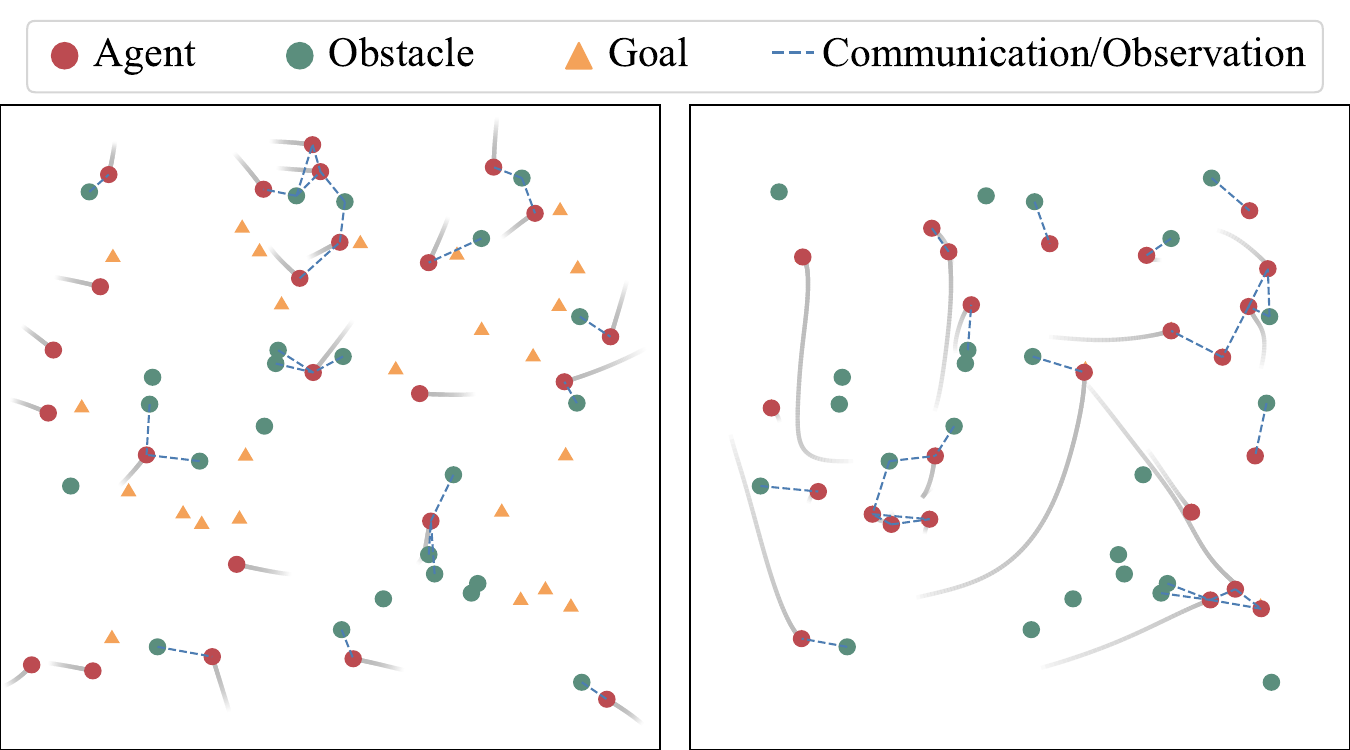}
		\caption{Cooperative Navigation scenario in Safe MPE, when the test episode (left) begins and (right) ends}
		\label{fig:24agents}
	\end{figure}
	
	\subsection{Supplementary Simulation Experiments}
	In addition to cooperative navigation, we also conduct experiments on  \textbf{Other Different Cooperative Tasks}, such as  Formation and Line. \par
	In the Formation task as described in \cite{EMP}, a scenario is set where $N$ agents are required to arrange themselves around a single landmark, which is positioned at the center of an $N$-sided regular polygon. The agents receive rewards at each time step based on their proximity to their designated positions. Additionally, they incur costs for colliding with one another. These positions are determined by solving a linear assignment problem at each time step, which takes into account the number of agents present in the environment and the desired radius of the polygon. For our experiments, we have set the target radius to a value of 0.5. \par
\begin{figure}[H]
		\centering
		\begin{subfigure}[b]{0.53\textwidth}
			\includegraphics[width=\textwidth]{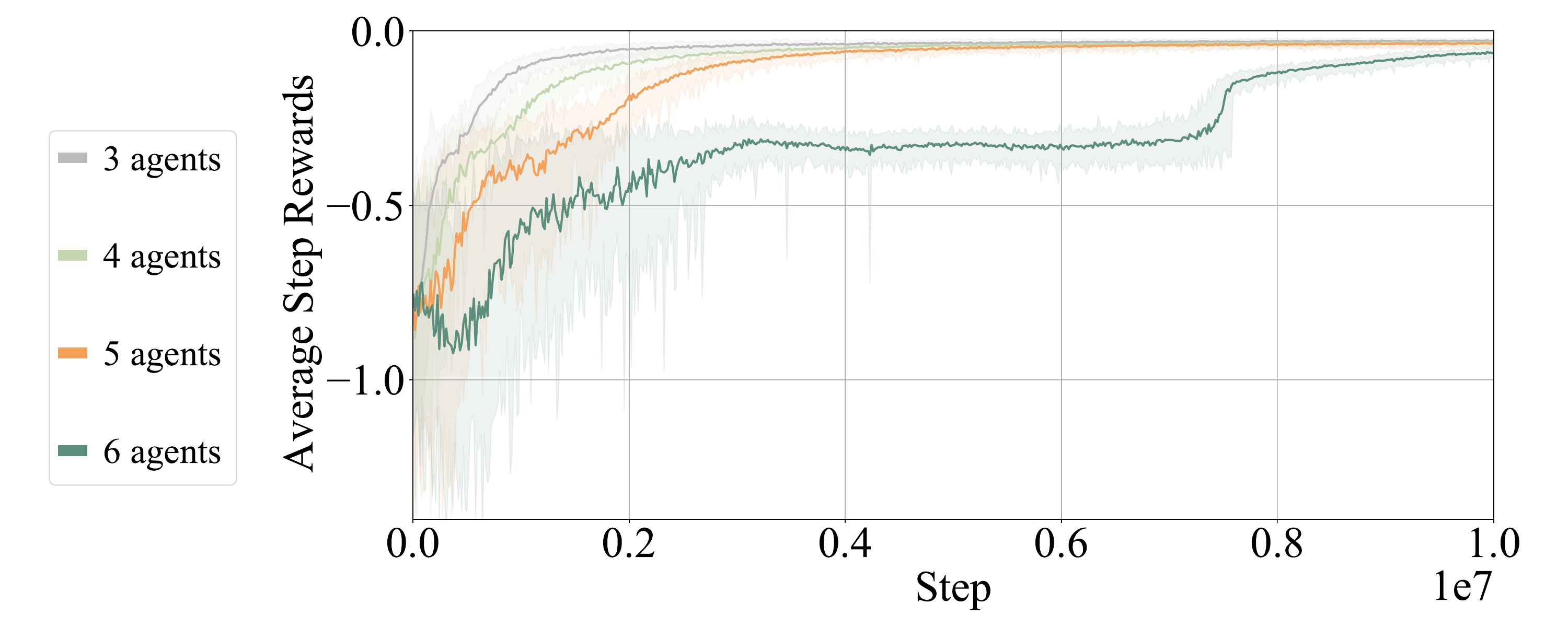}
			\caption{Average step reward per agent in Formation task}
			\label{fig:formation_reward}
		\end{subfigure}
		\begin{subfigure}[b]{0.46\textwidth}
			\includegraphics[width=\textwidth]{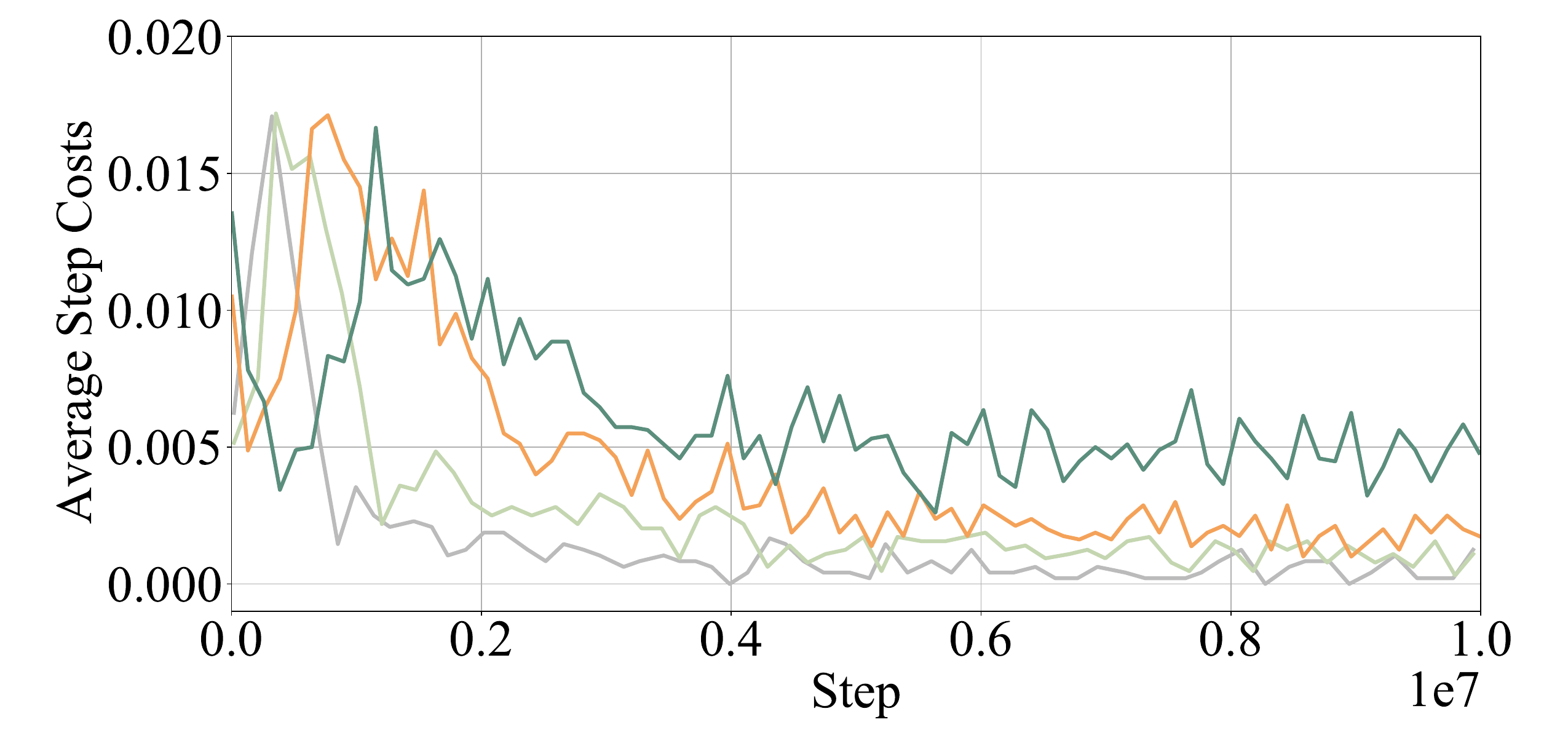}
			\caption{Average step cost per agent in Formation task}
			\label{fig:formation_cost}
		\end{subfigure}
		\begin{subfigure}[b]{0.53\textwidth}
			\includegraphics[width=\textwidth]{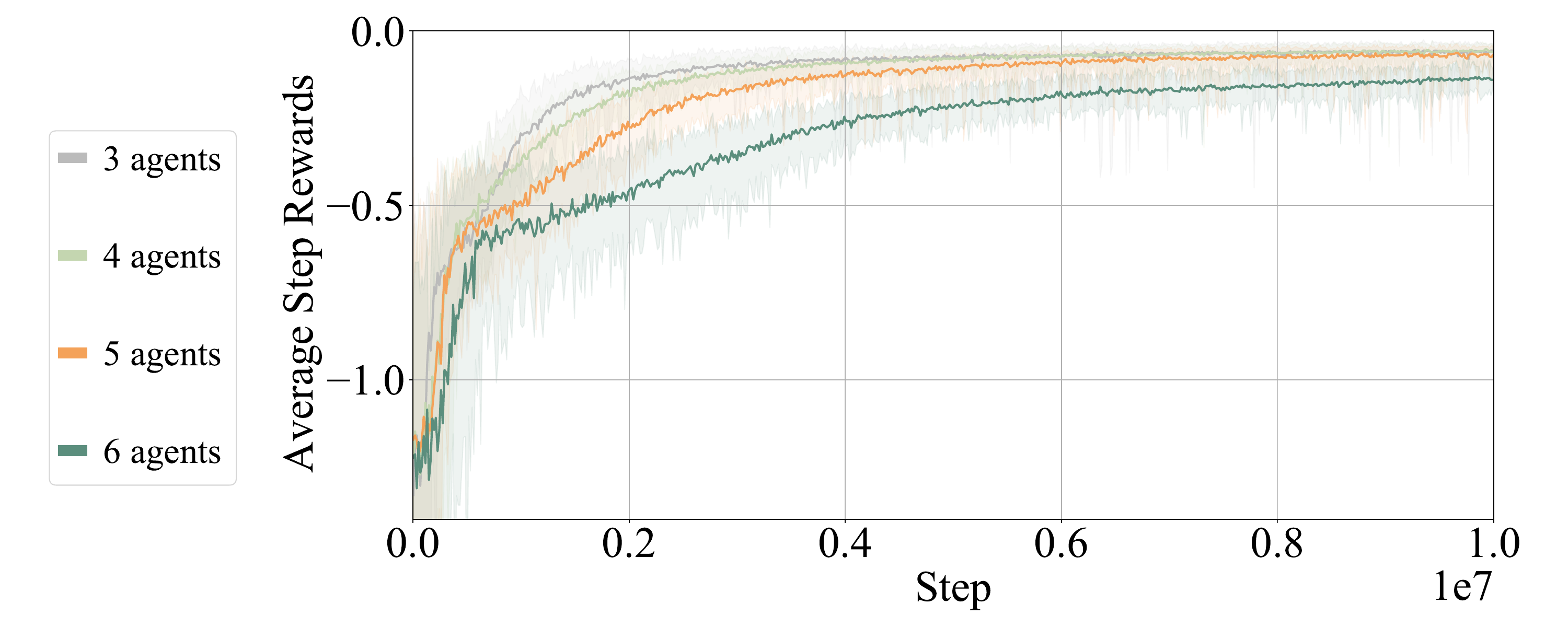}
			\caption{Average step reward per agent in Line task}
			\label{fig:line_reward}
		\end{subfigure}
		\begin{subfigure}[b]{0.46\textwidth}
			\includegraphics[width=\textwidth]{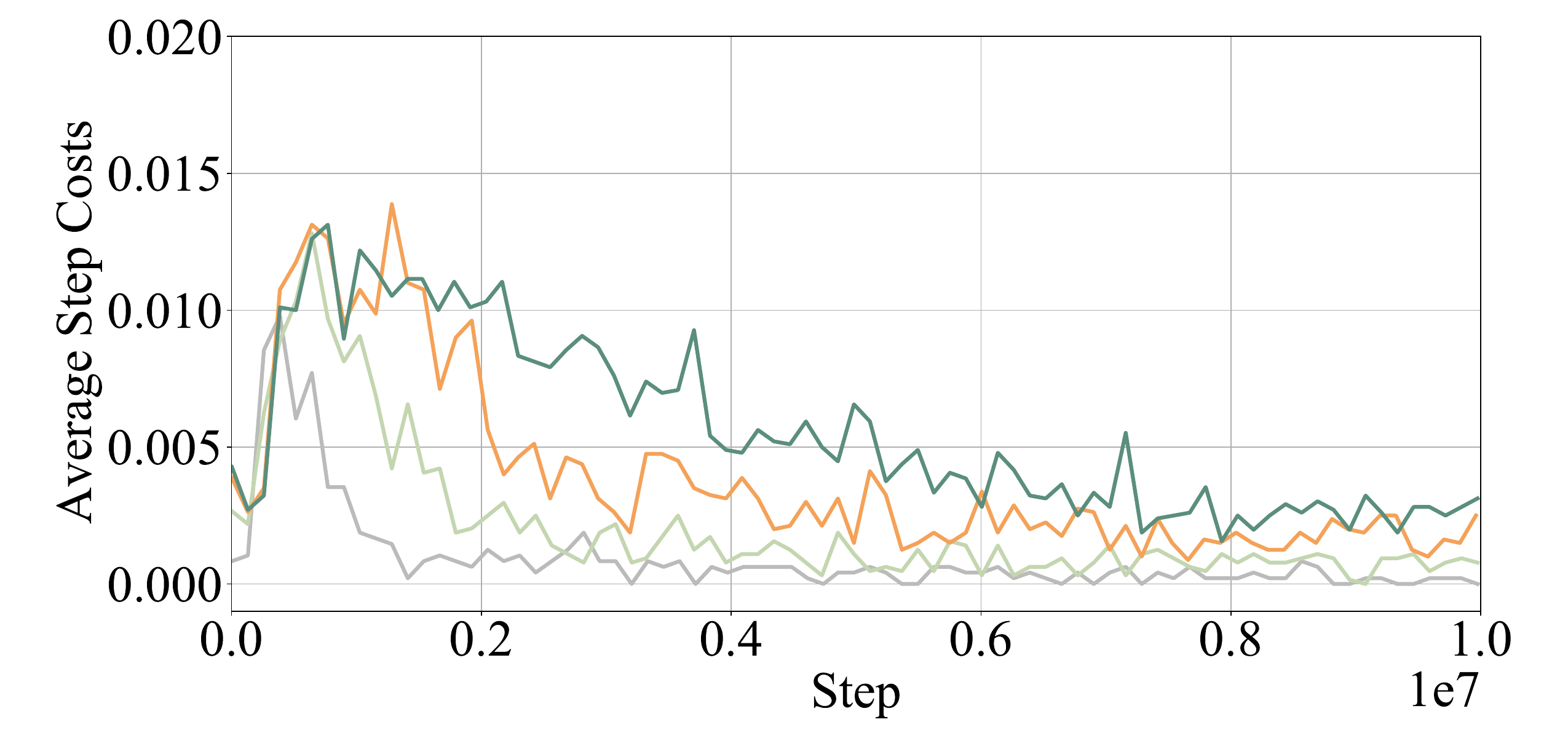}
			\caption{Average step cost per agent in Line task}
			\label{fig:line_cost}
		\end{subfigure}
		\label{different_tests}
		\caption{The training performance of SS-MARL on Formation and Line tasks. (a)(c) are average step rewards per agent during the training phase, (b)(d) are average step cost per agent during the training phase.}
	\end{figure}

	In the Line task outlined in the study \cite{EMP}, a setup is presented involving $N$ agents and a pair of landmarks. The objective for the agents is to disperse themselves evenly along a straight line that stretches between the two landmarks. As with the Formation environment, the agents earn rewards based on their nearness to the positions they are supposed to occupy. These positions are determined through the resolution of a linear sum assignment problem at every time step.
	\begin{figure}[H]
		\centering
		\begin{subfigure}[b]{0.2\textwidth}
			\fbox{\includegraphics[width=\textwidth]{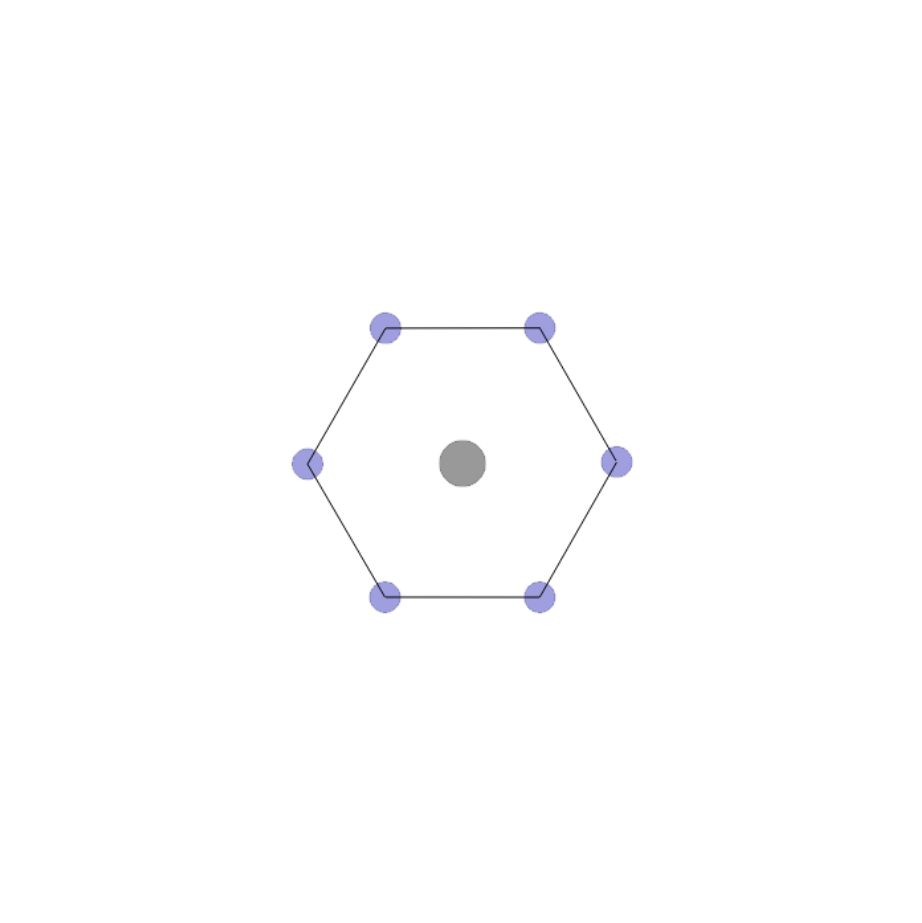}}
			\caption{Formation}
			\label{fig:formation}
		\end{subfigure}
		\hspace{10pt}
		\begin{subfigure}[b]{0.2\textwidth}
			\fbox{\includegraphics[width=\textwidth]{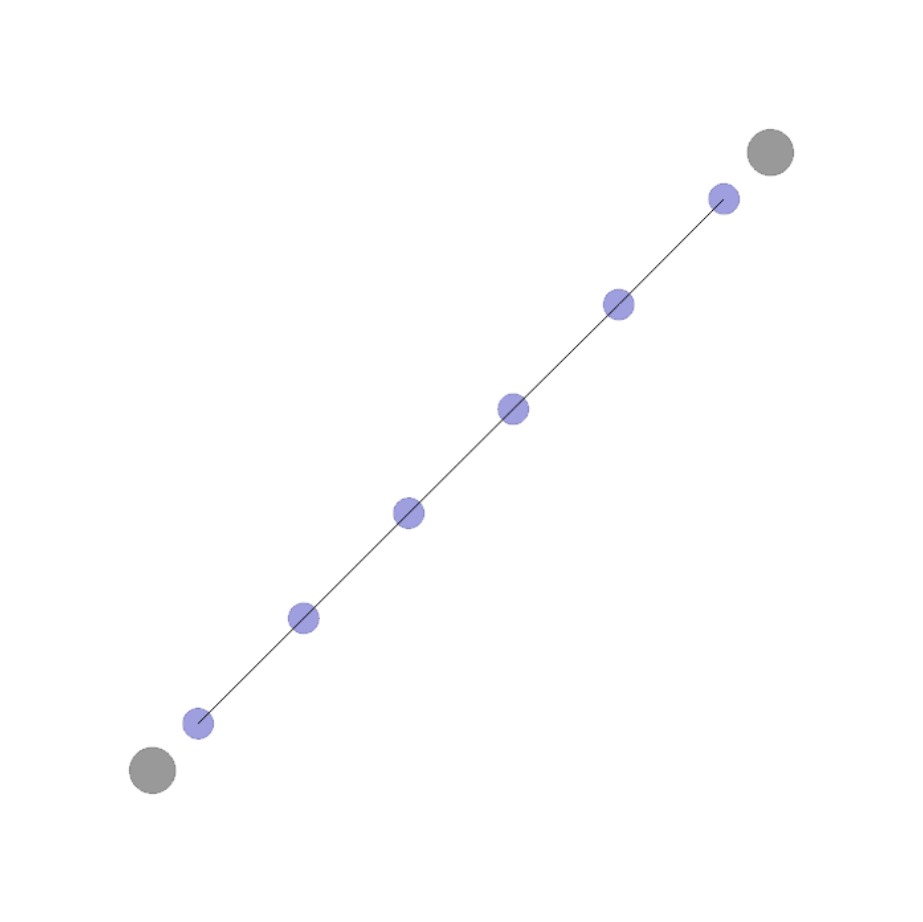}}
			\caption{Line}
			\label{fig:line}
		\end{subfigure}
		\caption{Formation task(a) and Line task(b), the blue circle is agent, the grey circle is landmark and the black lines between agents are communications}
	\end{figure}
	We trained SS-MARL on both Formation and Line tasks. The reward and cost curves during the training phase are shown in Figures \ref{fig:formation_reward},\ref{fig:formation_cost},\ref{fig:line_reward},\ref{fig:line_cost}. We set the cost constraint upper limit $c$ to 1, with the episode length set to 100, so the expected average step cost per agent is 0.01. From Figures \ref{fig:formation_cost} and \ref{fig:line_cost}, it can be seen that the costs in all scenarios are constrained within 0.01, and even if there are slight exceedances beyond 0.01, they can still be reduced through TRPO recovery steps to bring the constraint values back down. These sets of experiments demonstrate that SS-MARL not only performs well in cooperative navigation tasks but also shows good performance in other cooperative tasks. In the future, we will test the performance of the SS-MARL algorithm on more cooperative tasks, and we are also trying to transfer SS-MARL from the virtual environment to real-world robots.

	In order to further verify the \textbf{Zero-Shot Transfer Ability} of SS-MARL, we also tested the ability of models trained on the Cooperative Navigation task to transfer to these two tasks. We use a model trained on Cooperative Navigation scenarios with only $n=3$ to test in Formation and Line tasks with different number of agents. As with retraining, each agent in zero-shot transfer experiments is required to determine its own goal through a linear sum assignment problem.
	\begin{figure}[H]
		\centering
		\begin{subfigure}[b]{0.6\textwidth}
			\fbox{\includegraphics[width=\textwidth]{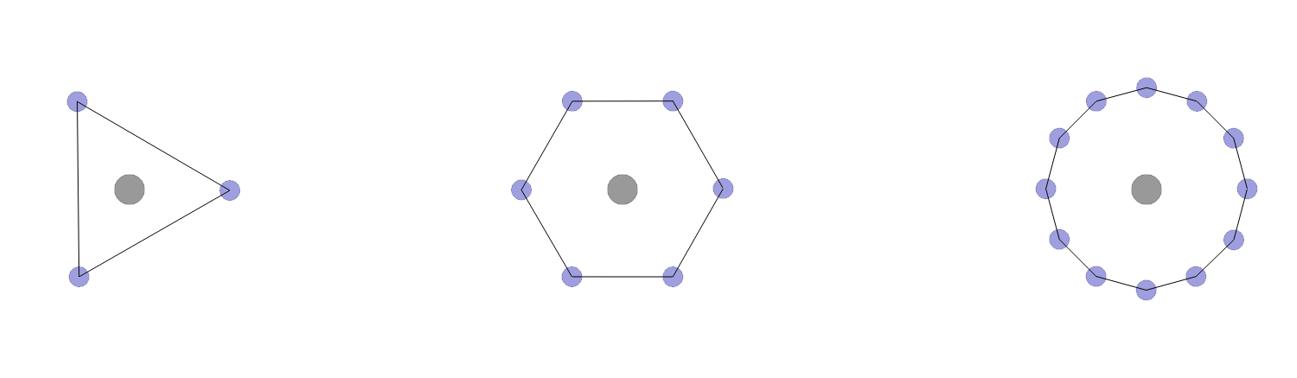}}
			\caption{Formation}
			\label{fig:formationt}
		\end{subfigure}
		\hspace{10pt}
		\begin{subfigure}[b]{0.6\textwidth}
			\fbox{\includegraphics[width=\textwidth]{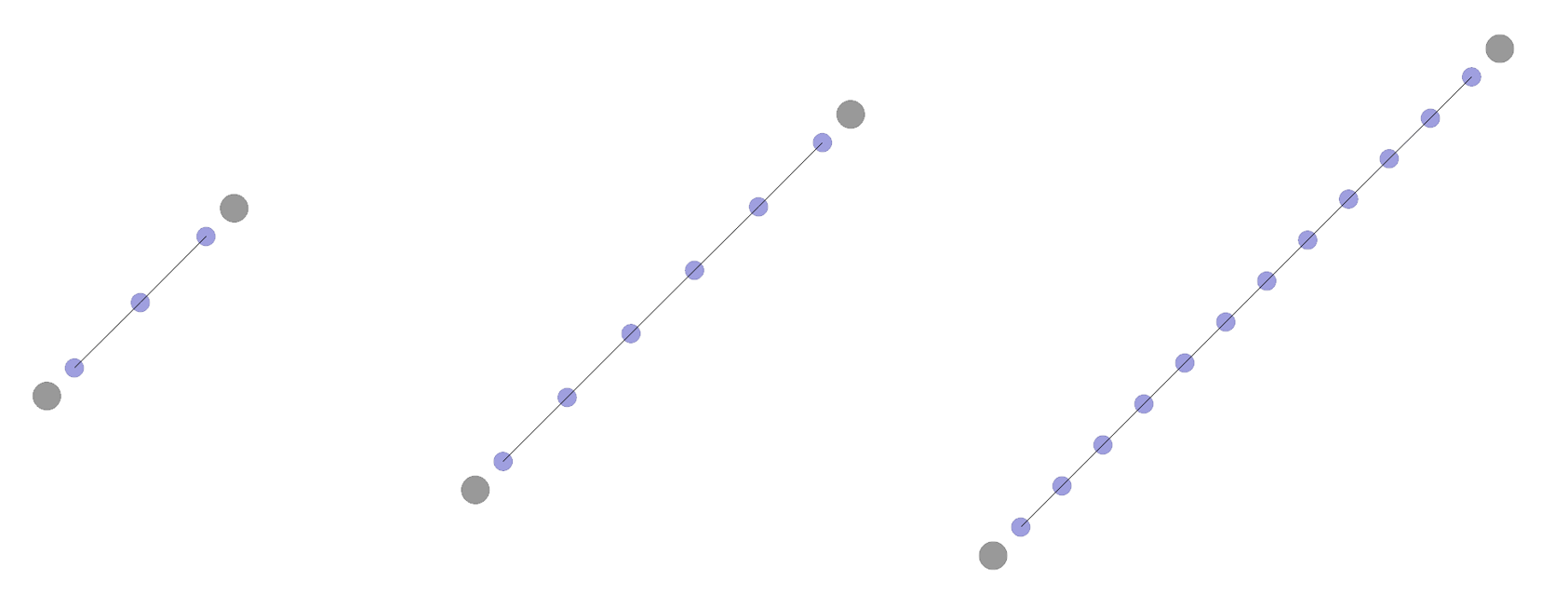}}
			\caption{Line}
			\label{fig:linet}
		\end{subfigure}
		\caption{Zero-shot transfer experiment on Formation task(a) and Line task(b), the blue circle is agent, the grey circle is landmark and the black lines between agents are communications. From left to right are 3,6, and 12 agents}
	\end{figure}
	The experimental results show that SS-MARL has great zero-shot transfer ability.
	\subsection{Supplementary Hardware Experiments}
\begin{figure}[H]
		\centering
		\includegraphics[width=0.3\textwidth]{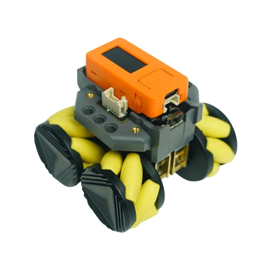}
		\caption{RoverC, a programmable and omnidirectional mobile miniature vehicle.}
		\label{fig:roverc}
	\end{figure}

	In addition to simulation experiments, we also conducted hardware experiments on the miniature vehicles\footnote{\href{https://docs.m5stack.com/en/hat/hat-roverc}{\emph{https://docs.m5stack.com/en/hat/hat-roverc}}} with Mecanum wheels shown as Figure \ref{fig:roverc}. We use the motion capture system to obtain the true position of vehicles. These true positions and the randomly generated obstacles will form an inherent graph structure. Then, the inheret graph structure is passed into the actor model which has been trained with SS-MARL(PS) in Safe MPE . The actions output by the actor are applied to the double integrator model, so we use dual-loop PID controllers for the low-level control of the miniature vehicles with Mecanum wheels.
	\par
\begin{figure}[H]
		\centering
		\begin{subfigure}[b]{0.6\textwidth}
			\includegraphics[width=\textwidth]{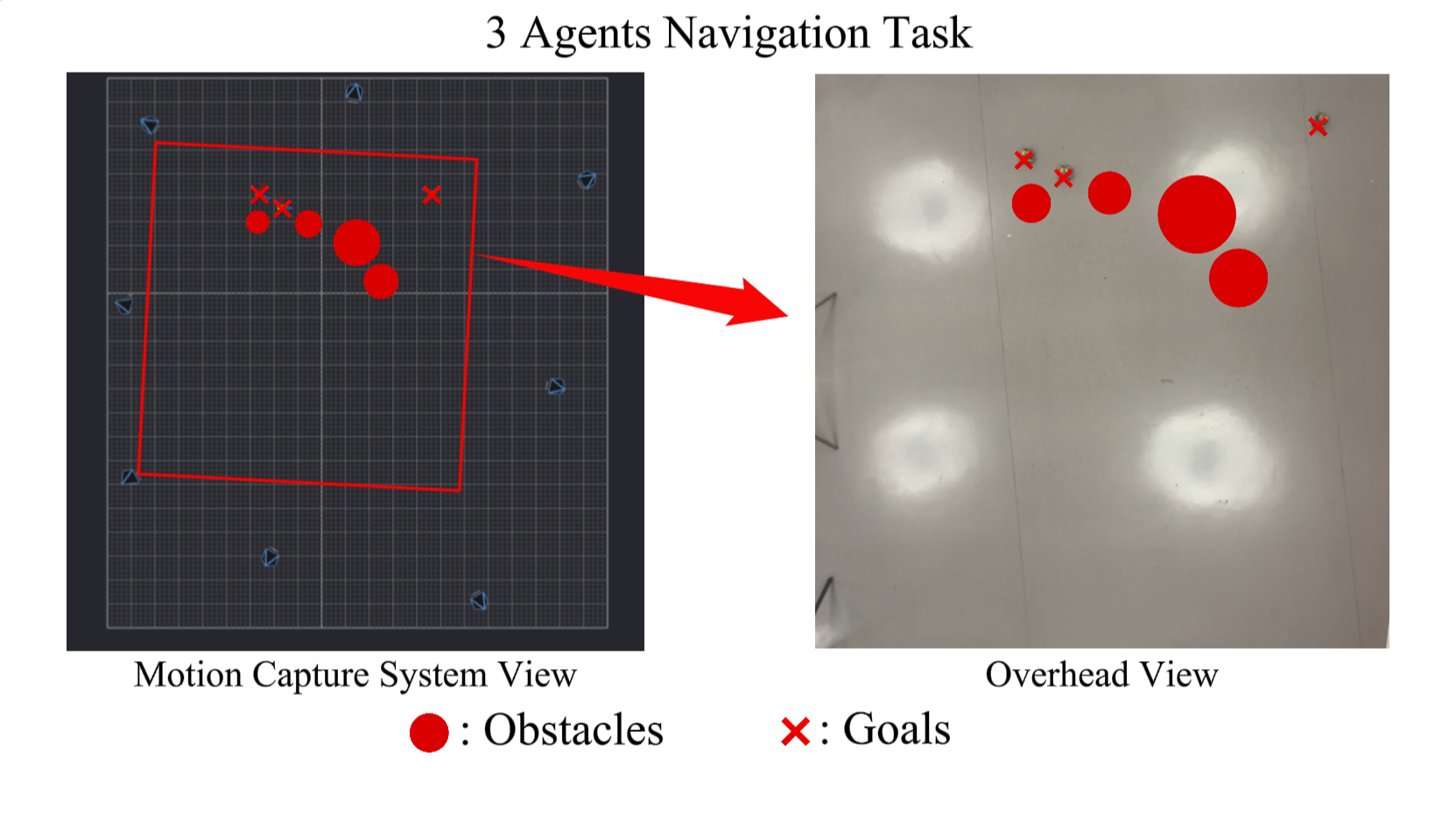}
			\caption{}
			\label{fig:navigation_3}
		\end{subfigure}
		\hspace{10pt}
		\begin{subfigure}[b]{0.3\textwidth}
			\includegraphics[width=\textwidth]{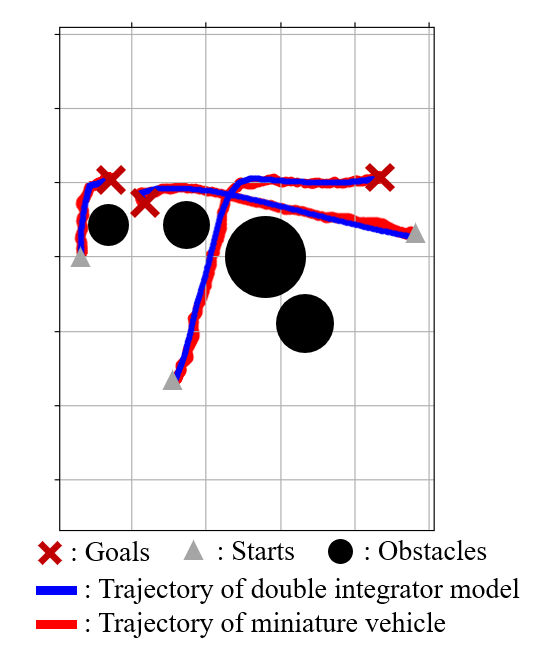}
			\caption{}
			\label{fig:navigation_3_line}
		\end{subfigure}
		\caption{Cooperative navigation task with 3 agents. The hardware implementation (a) and the trajectory (b).}
	\end{figure}
\begin{figure}[H]
		\centering
		\begin{subfigure}[b]{0.6\textwidth}
			\includegraphics[width=\textwidth]{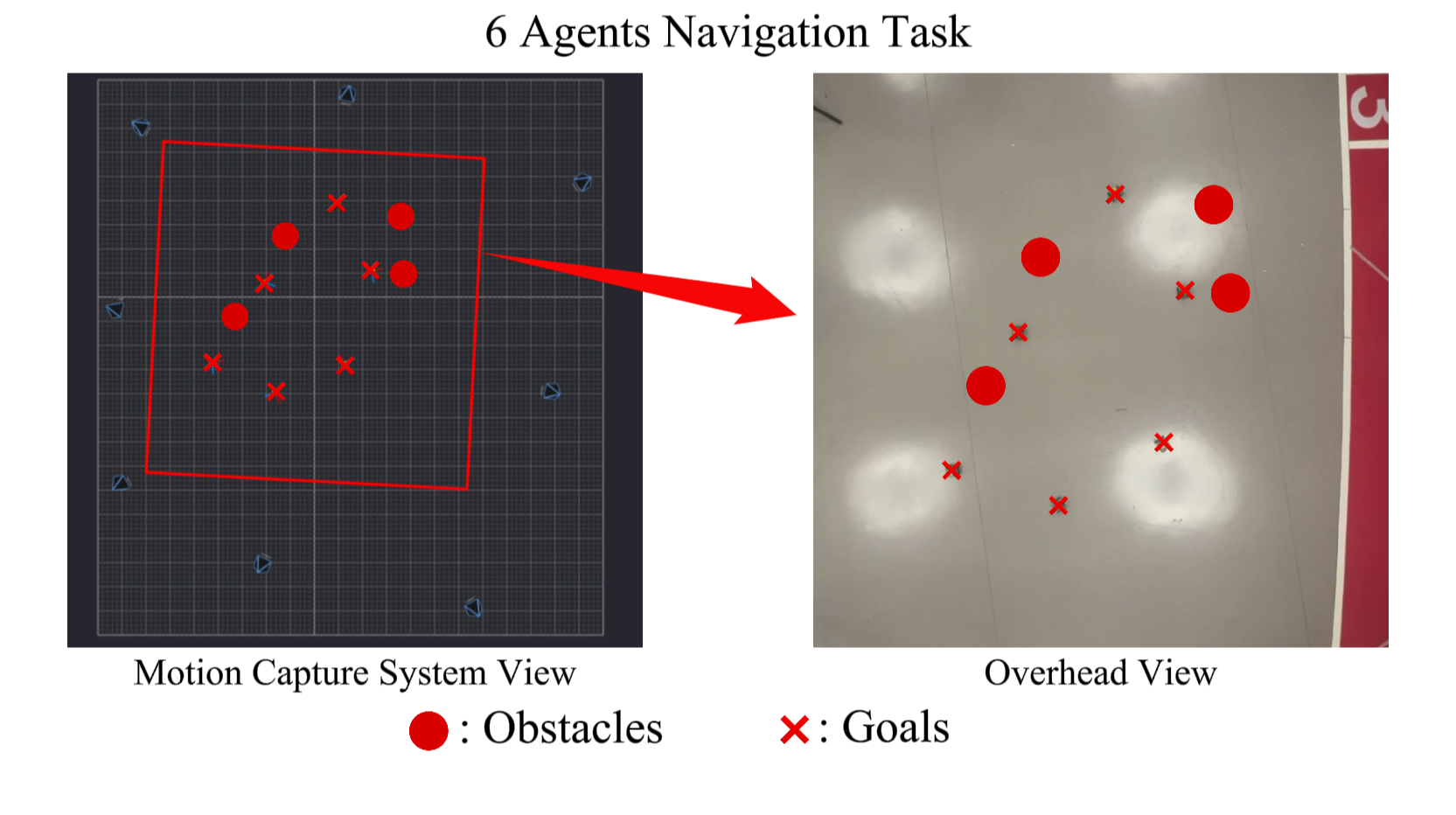}
			\caption{}
			\label{fig:navigation_6}
		\end{subfigure}
		\hspace{10pt}
		\begin{subfigure}[b]{0.3\textwidth}
			\includegraphics[width=\textwidth]{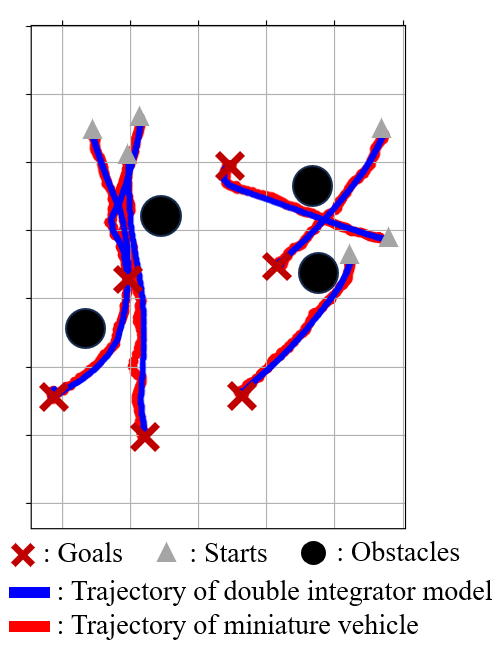}
			\caption{}
			\label{fig:navigation_6_line}
		\end{subfigure}
		\caption{Cooperative navigation task with 6 agents. The hardware implementation (a) and the trajectory (b).}
	\end{figure}

	We conducted hardware experiments on two cooperative navigation tasks. Figure \ref{fig:navigation_3}, \ref{fig:navigation_3_line}, \ref{fig:navigation_6} and \ref{fig:navigation_6_line} show the Motion Capture System View, Overhead View and the trajectories of both double inegrator model and miniature vehicle.
	
	\par
	The experimental results show that the trajectories of double integrator model are safe and no collision occurs. Under the action of the dual-loop PID controllers, the trajectories of miniature vehicles almost coincide with the trajectories of double integrator model. In these two experiments, the miniature vehicles do not collide with other vehicles or obstacles which are consistent with the simulation experiments in Safe MPE.
	\par
	These hardware experiments have shown us the potential of using SS-MARL in practical multi-robot applications. In the future, we will further conduct more experiments to explore the boundaries of the SS-MARL algorithm.
	\newpage
	\bibliographystyle{named}
	\bibliography{ijcai25}